\PassOptionsToPackage{table,xcdraw}{xcolor}
\documentclass[sigconf]{acmart}
\newcommand{\sol}{\textsc{ConceptLens}}
\newcounter{takeaway}

\setcopyright{acmlicensed}
\copyrightyear{2025}
\acmYear{2025}
\setcopyright{acmlicensed}\acmConference[CCS '25]{Proceedings of the 2025 ACM SIGSAC Conference on Computer and Communications Security}{October 13-17, 2025}{Taipei, Taiwan}
\acmBooktitle{Proceedings of the 2025 ACM SIGSAC Conference on Computer and Communications Security (CCS '25), October 13--17, 2025, Taipei, Taiwan}
\acmISBN{979-8-4007-1525-9/2025/10}
\acmDOI{10.1145/3719027.3744868}

\usepackage{enumitem}
\usepackage{xspace}
\usepackage{bm}
\usepackage{amsthm}
\usepackage[linesnumbered,ruled,vlined]{algorithm2e}
\usepackage{multirow}
\usepackage{multicol}
\usepackage{booktabs}
\usepackage{caption} 
\usepackage{etoolbox}
\usepackage[flushleft]{threeparttable}
\usepackage{array}
\usepackage{makecell}
\usepackage{mathtools}
\usepackage{soul} 
\usepackage{xcolor} 
\usepackage{caption} 
\usepackage{subcaption}
\usepackage{nicematrix}
\usepackage{listings}
\usepackage{float}
\usepackage{soul}

\usepackage{graphicx} 
\usepackage{caption}

\usepackage{url}

\usepackage{tcolorbox}
\tcbuselibrary{listings}
\tcbuselibrary{breakable}
\newcommand{\eg}{\textit{e.g.,}\xspace} 
\newcommand{\ie}{\textit{i.e.,}\xspace} 
\newcommand{\etal}{\textit{et al.}\xspace}

\newcommand{\one}{({\em i}\/)}
\newcommand{\two}{({\em ii}\/)}
\newcommand{\three}{({\em iii}\/)}

\usepackage{circledsteps}

\usepackage{framed}
\definecolor{formalshade}{rgb}{0.95,0.95,0.97}
\definecolor{darkblue}{rgb}{0.14,0.22,0.52}

\newenvironment{takeaway}{

\MakeFramed{\advance\hsize-\width\FrameRestore}}
{\endMakeFramed}

\newcommand{\tool}{\textsc{ConceptLens}\xspace}
\newcommand{\concept}{\textit{Concept Shift}\xspace}

\begin{document}

\title{What’s Pulling the Strings? Evaluating Integrity and Attribution in AI Training and Inference through Concept Shift}
\author{Jiamin Chang}
\affiliation{%
  \institution{University of New South Wales \& CSIRO's Data61}
  \city{Sydney}
   \country{Australia}
  }

\author{Haoyang Li}
\affiliation{%
  \institution{Macquarie University}
  \city{Sydney}
  \country{Australia}
  }

\author{Hammond Pearce}
\affiliation{%
  \institution{University of New South Wales}
  \city{Sydney}
  \country{Australia}
  }

\author{Ruoxi Sun}
\affiliation{%
  \institution{CSIRO’s Data61}
  \city{Adelaide}
   \country{Australia}
  }

\author{Bo Li}
\affiliation{%
  \institution{University of Illinois at Urbana–Champaign}
  \city{Champaign and Urbana}
  \country{United States}
  }

\author{Minhui Xue}
\affiliation{%
  \institution{CSIRO's Data61}
\city{Adelaide}
  \country{Australia}
  }

\renewcommand{\shortauthors}{Jiamin Chang et al.}



\begin{abstract}
The growing adoption of artificial intelligence (AI) has amplified concerns about trustworthiness, including integrity, privacy, robustness, and bias. To assess and attribute these threats, we propose \textsc{ConceptLens}, a generic framework that leverages pre-trained multimodal models to identify the root causes of integrity threats by analyzing \textit{Concept Shift} in probing samples. \textsc{ConceptLens} demonstrates strong detection performance for vanilla data poisoning attacks and uncovers vulnerabilities to bias injection, such as the generation of covert advertisements through malicious concept shifts. It identifies privacy risks in unaltered but high-risk samples, filters them before training, and provides insights into model weaknesses arising from incomplete or imbalanced training data. Additionally, at the model level, it attributes concepts that the target model is overly dependent on, identifies misleading concepts, and explains how disrupting key concepts negatively impacts the model. 
It uncovers sociological biases in generative content, revealing disparities across sociological contexts. 
\textsc{ConceptLens} reveals how otherwise safe training and inference data can be unintentionally and easily exploited to undermine safety alignment.

\end{abstract}

\maketitle
\section{Introduction}\label{sec_intro}

AI has emerged as a transformative technology, driving innovation across diverse domains such as healthcare, finance, autonomous systems, and creative industries~\cite{deng2014deep,papernot2018sok}. As these systems grow in complexity and prevalence, they also become prime targets for cyber-attacks, particularly those targeting the integrity of the data and models. Integrity ensures that the information processed and generated by AI systems remains accurate, consistent, and trustworthy under a wide range of scenarios~\cite{kaur2022trustworthy,liu2022trustworthy}. Compromising this integrity can lead to erroneous outputs, undermining the reliability of AI-driven decisions and actions. Key concerns include the risk of biased decision-making due to incomplete or skewed training data, susceptibility to adversarial attacks that exploit weaknesses in models, and the challenges of ensuring robust performance under unseen conditions. Extensive research has identified numerous trust-based risks, including security vulnerabilities in the face of adversarial perturbations~\cite{carlini2017adversarial,li2023sok,carlini2019evaluating}, privacy issues related to membership inference attacks~\cite{shokri2017membership,carlini2023extracting,zhang2020secret,carlini2022membership}, and the generation of biased or hateful content~\cite{qu2023unsafe}.  The importance of securing these AI systems is also reflected in recent legislative efforts -- for instance, in California, with SB 1047: Safe and Secure Innovation for Frontier Artificial Intelligence Models Act~\cite{salvador2024certified}. The bill's main provisions include mandatory pre-deployment safety assessments and robust cybersecurity measures for AI model developers, underscoring the need for trustworthy analysis pathways. Likewise, the EU AI Act~\cite{eprs2025algorithmic} seeks to address model performance bias by permitting the processing of ``special categories of personal data'' only under strict oversight as mandated by The General Data Protection Regulation (GDPR).

\begin{figure*}[t]
\centering
\includegraphics[width=1\linewidth]{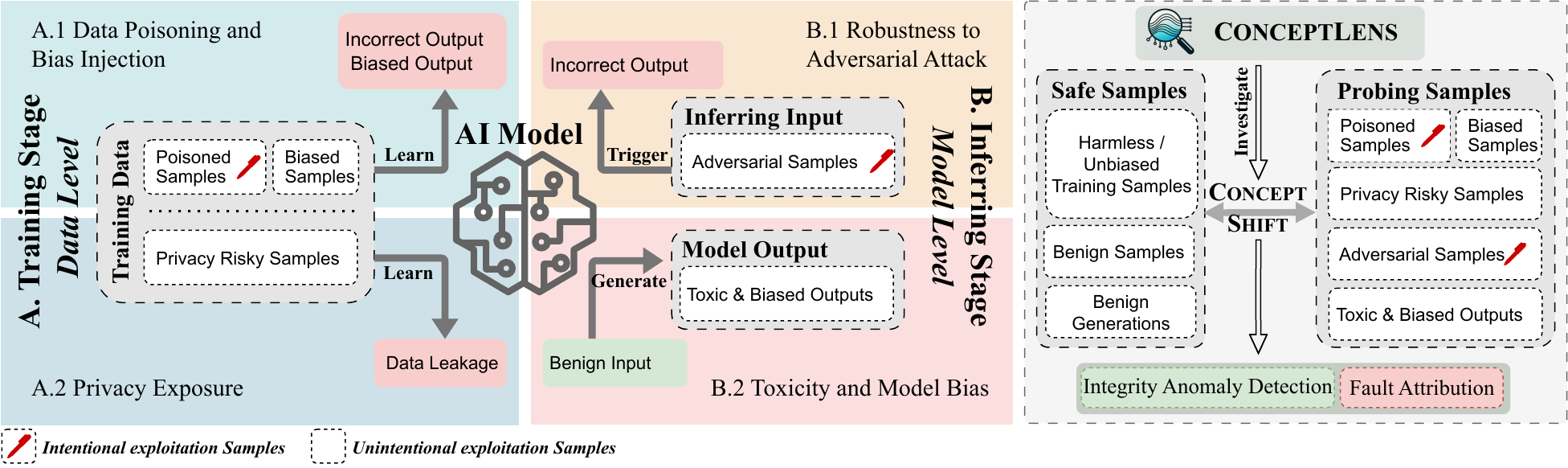}

\caption{An overview of the four trustworthy risks across both the training and inference stages is presented. Among these, only poisoned samples and adversarial samples are classified as intentional exploitation samples. \tool{} is designed to investigate concept shifts from safe samples to probing samples, enabling integrity anomaly detection and fault attribution. }
\label{fig:flow} 
\end{figure*}

Recent research has addressed areas such as adversarial machine learning, bias, and privacy preservation. In adversarial machine learning, methods have been developed to defend against adversarial examples designed to mislead models~\cite{sotgiu2020deep, ma2019nic, meng2017magnet, ma2018characterizing, shuo2024}. 
Research on model bias has produced metrics and strategies to mitigate the disproportionate impact of models on underrepresented groups~\cite{muennighoff2020vilio,sabat2019hate,zhu2020enhance,qu2023evolution,qu2023unsafe}. Privacy-preserving approaches, like federated learning and differential privacy, seek to safeguard sensitive information from leakage~\cite{zarifzadeh2024low,liu2024please}. However, these studies focus on specific aspects like robustness or fairness and rely on heuristics for protections, lacking a unified framework for understanding how errors or biases propagate during training and inference. Existing tools mainly target adversarial faults in AI models~\cite{shuo2024}, but there are few systematic methods to evaluate and attribute integrity threats. Recently, He~\etal~\cite{he2024what} found that benign data can degrade model safety after fine-tuning.  Multimodal models where modalities interact further increases the risk of unintended interactions leading to susceptibility to adversarial inputs, biased outputs, and privacy leakage. 
There is a clear need for a comprehensive solution to these challenges. Our work addresses this through the introduction of
\tool{}, a framework for evaluating integrity threats at both the data and model levels by utilizing \concept{}. 

In this study, we define a \textit{concept} as an abstract and semantic representation of a characteristic or feature. Artificial Intelligence can be viewed as an information processing pipeline with two primary information flows~\cite{wu2024system,sun2024unifying}: \one~from training data to model during the training phase, and \two~from inputs to outputs during the inference phase. A trustworthy model should ensure that the information flowing from inputs (either training or testing inputs) to model outputs is consistent (i.e. does not \textit{shift}). For example, in a multimodal text-to-image model, the generated image should align with the input textual description, at least at a conceptual level~\cite{iso24028}. Therefore, we argue that, to ensure the integrity and trustworthiness of AI models, \concept{} should be avoided and carefully measured. (Here, we note that \concept{} is distinct from Concept Drift~\cite{barbero2022transcending}, which refers to changes over time in the statistical properties of the target variable or data distribution.) 

During training, models learn patterns and representations from large datasets. This process involves extracting, transforming, and encoding information into the model's structures and parameters. However, this flow is vulnerable to data-level integrity issues. For instance, biased or maliciously manipulated training samples can inject incorrect information, resulting in models that internalize errors or harmful biases. During inference, models process input samples (\eg testing data) by combining information from the input samples with the knowledge stored in the model during training. The outputs depend on both the quality of the input and the integrity of the learned representations (\eg learned concepts). Here, model-level integrity threats arise, such as adversarial attacks that manipulate inputs to exploit learning errors and mislead the model into generating incorrect predictions.

However, achieving and measuring this consistency is nontrivial, as real-world data often introduces unforeseen complexities. In our study, based on the measurement of \concept{}, we propose \tool{} as a framework designed to capture the consistency (or shift thereof) in the information flows during model training and inference. \tool{} operates based on three key features derived through a combination of coarse- and fine-grained alignment techniques: \one~\textit{vision \& concept linear abstract feature similarity} that measures the alignment between visual and conceptual representations learned by the model, ensuring that the model's high-level abstractions are consistent with input data; \two~\textit{concept prediction posteriors} that examines the reliability and strength of predictions for specific concepts; and \three~\textit{attention localization} that evaluates the positional importance of specific concepts in the input, providing interpretability and identifying potential sources of inconsistency or misalignment in model predictions. 

The key contributions of this paper are as follows:

\begin{itemize}
\item \textbf{Concept Shift.} We define and propose \concept{} as a systematic technology for understanding and addressing integrity issues in AI training and inference processes, providing mechanisms for detecting, analyzing, and mitigating integrity risks that arise from changes in conceptual understanding. It facilitates the identification of subtle disruptions that affect model performance -- often overlooked in traditional integrity assessments.
\item \textbf{\tool{}.} Based on \concept{}, we propose \textsc{Concept\-Lens}, a comprehensive framework for evaluating model trustworthiness by extracting features from both coarse-grained alignment (which establishes vision and concept features) and fine-grained alignment (which generates concept prediction posteriors and attention localization). Unlike existing integrity evaluation studies, \tool{} not only addresses \textit{intentional} integrity risks, such as data poisoning and adversarial attacks, but also identifies \textit{unintentional} integrity threats in benign samples (these include, but are not limited to, bias injection during training, privacy exposure risks, and bias in model outputs).
\begin{itemize}
\item \textit{Proactive detection}, such as detection of privacy exposure, bias injection, and adversarial detection, can be used to improve data collection practices to minimize the inclusion of such samples in the first place.
\item \textit{Reactive detection}, such as bias output detection, identifies and quantifies sociological bias in model outputs, offering insights into how such biases can be reflected and propagated.
\end{itemize}

Through detection, we can evaluate the vulnerability of different models to these existing threats and identify issues within the models. Though not a defense in itself, being able to detect such issues is important for deriving suitable defenses downstream.
\item \textbf{Evaluation of Integrity.} We conduct extensive evaluations of integrity at both the data and model levels. At the data level, we assess threats such as bias injection and privacy exposure risks. At the model level, we examine adversarial robustness and model output bias. These evaluations span several mainstream models, including both single-modal and multimodal.
\item \textbf{Fault Attribution.} We develop methods to systematically attribute the causes of integrity failures, offering insights into model weaknesses stemming from incomplete or imbalanced training data. Our approach identifies concepts that the target model is overly dependent on, highlights misleading concepts, and explains how disrupting key concepts negatively affects model performance.
\end{itemize}

This paper defines and explores methodologies for evaluating integrity and attribution at both the data and model levels. We hope that our study offers actionable insights to advance the trust of AI systems, particularly in the context of complex multimodal applications. The code and artifacts are made publicly available at: \url{https://github.com/trust-in-ai/conceptlens}. 

\section{Background and Related Work}\label{sec_background}

In this section, we introduce recent studies related to AI integrity and trustworthiness, with a focus on data and model integrity.

\subsection{Data Integrity}

\noindent \textbf{Data poisoning and bias injection.}
A data poisoning attack involves injecting malicious data into the training set to disrupt the learning process, aiming to degrade model performance or manipulate its behavior to align with the attacker’s objectives~\cite{MM2023Zhai,Oakland2024Shan}. Detection has been a mainstream approach to preventing backdoor and poisoning attacks~\cite{TMLR2022Lu, Oakland2024Shan}; however, the detection performance achieved by existing methods has generally been low, as shown in Table~\ref{tab_defense_mislearning}. To address this, \tool{} leverages additional features through multimodal alignment to improve detection performance.

Toxic and harmful content increases with the expansion of training datasets, Birhane~\etal~\cite{birhane2023hate} show that hateful and racist outputs from models tend to increase when utilizing larger state-of-the-art open-source datasets. Building on this, we further measured the scenario of advertising generation in text-to-image models, finding that samples containing advertisements do not disrupt image-text alignment but still pose risks. We believe that solutions for this unintentional exploitation remain open, consistent with the perspective of the recent work by He~\etal~\cite{he2024what}.

\noindent \textbf{Privacy exposure.}
When neural networks are trained on sensitive datasets (\eg medical data), it is essential to ensure that the trained models are privacy-preserving. However, membership inference attacks~\cite{shokri2017membership,carlini2023extracting,zhang2020secret,carlini2022membership,salem2018ml,salem2020updates} can allow attackers to determine if inputs were in the training data distribution. This is usually done by analysis of the posterior probabilities, which are the raw output scores produced by a shadow model trained by samples with the same distribution as the training dataset~\cite{shokri2017membership,salem2018ml,liu2022membership,li2021membership,jia2019memguard}. For instance, datasets in the medical field often contain private information, and the disclosure of such datasets can lead to privacy concerns. Sample hardness, first proposed by Carlini~\etal~\cite{carlini2022membership}, is reflected by posterior differences between in- and out-models to stress that the membership of some samples in the dataset is easier to be inferred by that of the others. However, subsequent works still analyze membership based on posteriors~\cite{zarifzadeh2024low,liu2024please}. Therefore, a more comprehensive understanding about sample differences in membership information should be sought. In this paper, we harness \tool{} as an independent black-box assessment tool for analyzing membership inference vulnerabilities. It surpasses LiRA~\cite{carlini2022membership}, which demands white-box access and the training of 64 shadow models.

\subsection{Model Integrity}

\noindent \textbf{Adversarial perturbations.}
Deep Neural Networks (DNNs) have been adopted for classification tasks in various applications, including highly security-sensitive areas such as face recognition. However, malicious adversaries can manipulate classification models to produce desired outputs using carefully crafted inputs~\cite{goodfellow2014explaining,madry2017towards,moosavi2016deepfool,papernot2016limitations,carlini2017towards,kotyan2022adversarial}. These samples are created by adding very small perturbations (changes) to legitimate inputs and used to cause models to misclassify. To better protect models, existing works have designed detection-based methods to identify adversarial examples for image classification tasks~\cite{sotgiu2020deep,ma2019nic,meng2017magnet,ma2018characterizing,shuo2024}. To understand why models make specific errors using concepts, previous works~\cite{kim2018interpretability,bau2020understanding} offer valuable tools for neural network interpretation, focusing on concept-based explanations, neuron-level semantics, and feature importance, yet they fall short in explaining conceptual faults. The most recent work~\cite{shuo2024} is designed to diagnose various types of model faults by interpreting latent concepts, which relies on mapping high-dimensional input to a low-dimensional latent space. 

Multimodal Vision-Language pre-training models (VLPs) have demonstrated considerable capabilities across a range of Vision-Language tasks~\cite{li2021align,dou2022empirical,radford2021learning}. 
However, as with image classification models, adversaries can manipulate VLP models~\cite{zhang2022towards,lu2023set} with adversarial samples to impact outputs. However, there is currently no targeted mitigation to defend against this kind of multimodal attack. 

Unlike current proposed methods, which are limited to single-modality tasks and abstract concept-level detection, \tool{} handles multimodal attacks and delivers human-interpretable explanations for model failures. By integrating text-modality cross-attention maps with Grad-CAM~\cite{selvaraju2020grad}, we extend interpretability beyond unimodal classification tasks and vision-language pretraining tasks.

\noindent \textbf{Toxic and biased generation.}
Text-to-image models are prone to generating unsafe images, raising concerns about the use of AI-Generated Content (AIGC) in contexts such as front-facing business websites or direct consumer communications. Wu~\etal~\cite{wu2024image} quantitatively assessed the safety of model-generated images, evaluating whether they contain factors such as violence, gore, or explicit content. 
 
Previous studies~\cite{muennighoff2020vilio,sabat2019hate,zhu2020enhance} focus on meme detection using multimodal frameworks, including work initiated by Facebook's Meme Challenge~\cite{kiela2021hateful}. Qu~\etal~\cite{qu2023evolution,qu2023unsafe} recently explored meme evolution and AI-generated  meme variants in multimodal models.
However, whether the generated meme variants contain sociological bias (\eg a ``pepe the frog'' with a specific country flag or traditional cultural element) has not been discussed. 
In this study, we particularly focuses on such specific type of unsafe image generation: hateful memes with sociological blending, which are subtle and difficult to detect. Specifically, we study toxic AI models to automatically quantify AI trust by measuring biased generation.

\section{\tool}\label{sec_methodology}

In this section, we first introduce \concept{} and then propose \tool{}, a framework for evaluating the integrity of AI models. The framework assesses their trustworthiness during both the training and inference stages through explainable concepts.

\subsection{Concept Shift}
\label{sec:concept_shift}
Any given sample, such as an image of a kitten, can be associated with numerous semantic concepts -- such as the kitten’s brown eyes, pointed ears, and white fur. 
Here, we define a concept as an abstract and semantic representation of a characteristic or feature. Language acts as a medium for humans to aggregate and express these concepts, facilitating the description of multimodal information and the explanation of language itself. In AI, a trustworthy model should ensure that the information (\eg concepts) flowing from inputs to outputs remains \textit{consistent}. For example, an image generated based on the description ``a kitten with brown eyes, pointed ears, and white fur'' should accurately reflect these characteristics. Similarly, a caption generated from such an image should describe these same features. Any shift or misalignment in these concepts between inputs and outputs, such as altering the textual description or introducing noise to the image, can adversely affect the model's learning or inference process.

Unfortunately, the concepts ``understood'' or ``learned'' by AI models can often differ significantly from a human perspective, as AI models rely on mathematical and statistical features rather than intuitive understanding. This discrepancy makes it possible for concept shifts to be injected or manipulated with malicious intent in a subtle and stealthy manner, such as through changes in the latent space, which are difficult for humans to recognize. For data level, during the learning phase, disruptions of data integrity, such as data poisoning or biased training, may cause the concepts of certain samples to shift incorrectly, leading to erroneous knowledge acquisition. In addition, with unique concepts, models may become overly reliant on them, resulting in excessive memorization and potential privacy leaks. For model level, during the inference phase, attackers can manipulate the concepts of input images to undermine the model’s integrity, such as through adversarial samples that exploit fundamental errors in learning, leading to incorrect predictions. Furthermore, models may generate biased outputs that reflect these manipulated concepts. At any stage, concept shifts are fundamentally tied to a model’s trustworthiness.

To address these challenges and threats to AI integrity, we designed \tool{} using a vision-language pre-trained model to extract the semantic concepts of samples and measure shifts in these concepts, enabling the evaluation and mitigation of such risks (as shown in Figure~\ref{fig:flow}). The deployment of AI models can be broadly divided into the training phase and the inference phase. In both phases, we measure concept shifts in samples using \tool{} to detect anomalous samples, filter them out, and attribute the root causes of these anomalies. In the training phase, the model relies on training data to learn, making it susceptible to data-level trustworthiness risks. We detect and exclude suspicious poisoned samples or those with privacy risks from the training set to ensure the model learns from reliable data. In the inference phase, the most critical issue is malicious users employing adversarial samples to manipulate the model's predictions. Additionally, the model might generate outputs containing malicious or biased content. These anomalous samples should also be intercepted to maintain the model trust.

\subsection{Design of \tool{}}

\begin{figure}[t]
\centering
\includegraphics[width=1\linewidth]{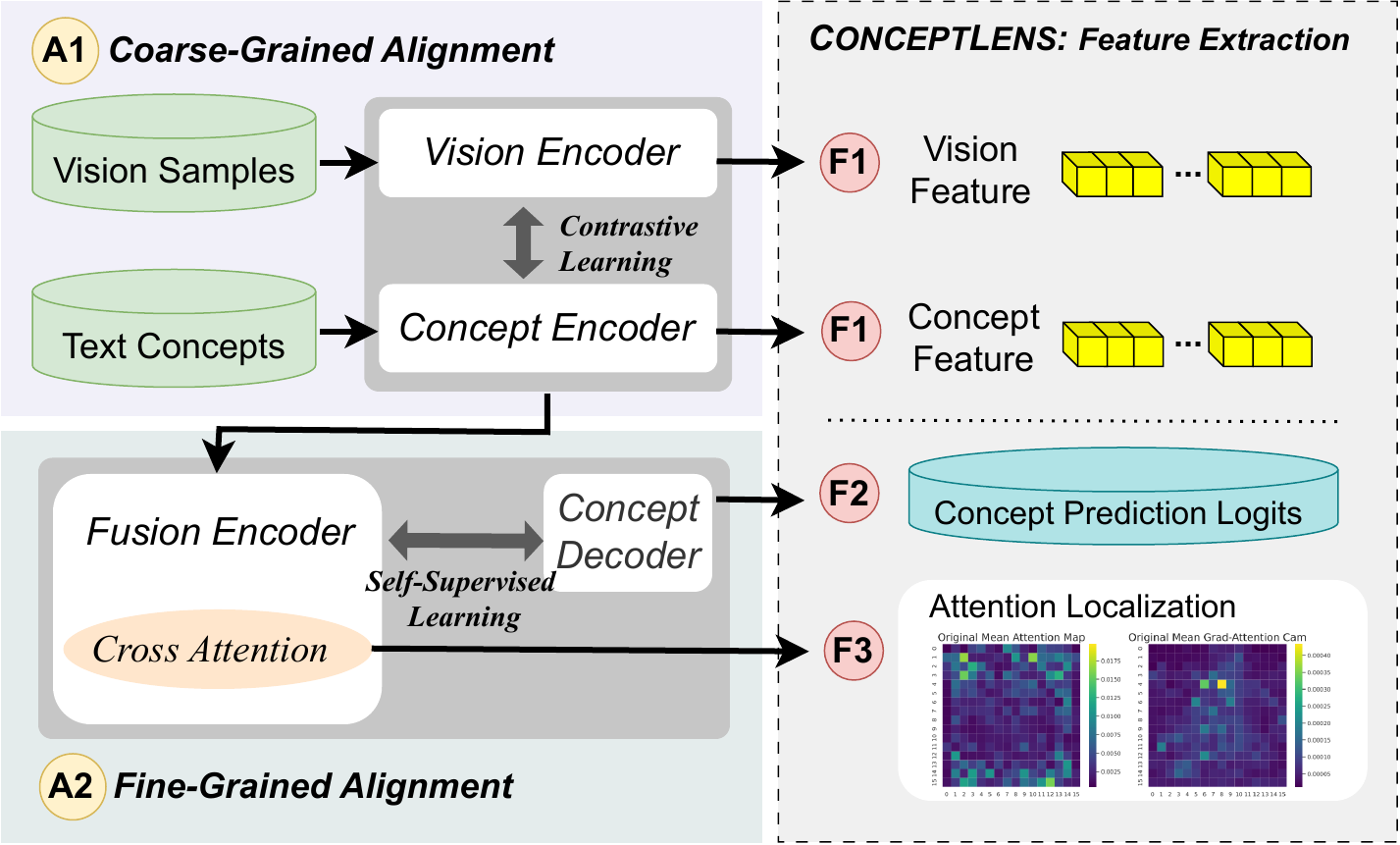}
\caption{An overview of \tool{}. It begins by \Circled{A1} establishing Vision and Concept Features \Circled{F1}, then progresses to forming concept prediction posteriors \Circled{F2} and attention localization \Circled{F3} through fine-grained alignment \Circled{A2}. \tool{} leverages probing samples to integrate detection protection and attribute model weaknesses.}
\label{fig:overview} 
\vspace{-1em}
\end{figure}

The framework of \tool{} is depicted in Figure~\ref{fig:overview}. To start an analysis, vision samples are provided by the users along with customized concept segments (\ie text concepts). For example, for an input image of an airplane, the concept segment could be: ``an image of an airplane with its wings and body in the sky''. For image samples, the label can be regarded as a concept, while for multimodal samples, the image caption itself contains a wealth of concepts. The framework then aligns visual features with conceptual features to provide explainable conceptual representations through multimodal conceptual space alignment and feature extraction. 

\subsubsection{Multimodal Conceptual Space Alignment}
Leveraging VLP model ALBEF~\cite{li2021align}, the framework contains a vision encoder, a concept (language) encoder for coarse-grained alignment between two modalities, and a multimodal fusion encoder for further fine-grained mapping. We select ALBEF for this study due to its open-source availability and pre-training on large-scale datasets. However, our framework is model-agnostic and can be seamlessly applied to any model that provides generalized image and text encoders alongside a fusion encoder. 

\noindent\textbf{\Circled{A1}: Coarse-gained alignment by contrastive learning.}
The vision encoder is a 12-layer ViT-B/16~\cite{dosovitskiy2020image}.  An input image $I$ is encoded into a sequence of embeddings: $\{v_{\text{cls}}, v_1, \ldots, v_N\}$, where $v_{\text{cls}}$ represents the embedding of the [CLS] token. The text encoder is initialized with the first 6 layers of the BERT model~\cite{kenton2019bert}, and converts an input concept text $T$ into a sequence of embeddings $\{w_{\text{cls}}, w_1, \ldots, w_N\}$.  Image-Text Contrastive Learning is used to better align unimodal representations from the vision and concept encoders by using a similarity function $s = g_v(v_{\text{cls}})^\top g_w(w_{\text{cls}})$ to calculate the feature contrastive loss, where $g_v$ and $g_w$ map the image and text [CLS] embeddings to lower-dimensional representations, respectively. The feature contrastive loss is defined as:
\begin{equation}\footnotesize
\mathcal{L}_{\text{feature}} = \frac{1}{2} \mathbb{E}_{(I,T) \sim \mathcal{D}} \left[\mathcal{H}(y^{i2t}(I), p^{i2t}(I)) + \mathcal{H}(y^{t2i}(T), p^{t2i}(T))\right], 
\end{equation}
where $\mathcal{H}$ denotes the cross-entropy loss, $y^{i2t}(I)$ and $y^{t2i}(T)$ are the ground-truth one-hot similarity labels for image-to-text (i2t) and text-to-image (t2i) predictions, and $p^{i2t}(I)$ and $p^{t2i}(T)$ represent the predicted similarity scores. The expectation $\mathbb{E}_{(I,T) \sim \mathcal{D}}$ is taken over the data distribution $\mathcal{D}$. 

\noindent\textbf{\Circled{A2}: Fine-grained alignment by self-supervised learning.}
The multimodal fusion encoder is a 6-layer transformer, where it is initialized with the last 6 layers of {the BERT model~\cite{kenton2019bert}} with image embedding and text embedding as inputs. 
A Masked Language Modeling task is designed to guide the integration using self-supervised learning, which utilizes both the image and the contextual text to predict the masked words with a one layer decoder. The goal of Masked Language Modeling (mlm) loss is to minimize the cross-entropy:
\begin{equation}\footnotesize
\mathcal{L}_{\text{mlm}} = \mathbb{E}_{(I, \hat{T}) \sim \mathcal{D}} \left[\mathcal{H}(\mathbf{y}^{\text{msk}}, p^{\text{msk}}(I, \hat{T}))\right], 
\end{equation}
where $\hat{T}$ denotes a masked text, $p^{\text{msk}}(I, \hat{T})$ denotes the model’s predicted probability for a masked token, and $\mathbf{y}^{\text{msk}}$ is a one-hot vocabulary distribution where the ground-truth token has a probability of 1. The image features are integrated with the text features through cross attention at each layer of the multimodal encoder. By recovering obscured words, the multimodal fusion encoder gains the ability to capture low-level features, providing fine-grained local information that enables the model to recognize and align subtle details in multimodal data, so that it can match specific words to corresponding visual objects.

\noindent\textbf{Datasets for pre-training.}
The model is pre-trained on a dataset comprising 14.1 million images is used for model pretraining, sourced from Conceptual Caption~\cite{sharma2018conceptual}, SBU Captions~\cite{ordonez2011im2text}, MS COCO~\cite{lin2014microsoft}, Visual Genome~\cite{krishna2017visual}, and Conceptual 12M~\cite{changpinyo2021conceptual}, following the original settings of ALBEF~\cite{li2021align}. We do not fine-tune the model further nor clean their data. The pre-trained model is readily available from open or commercial repositories.

\subsubsection{Feature Extraction} 
\label{sec:feature_extract}
\noindent \textbf{\Circled{F1}: Vision \& concept linear abstract feature similarity.} From the vision and language encoders, we get multimodal abstractly aligned features $v_{\text{cls}}$ and $w_{\text{cls}}$ (the 2 lower-dimensional (256-dimensional) abstract contextual information representations). Their similarity score from the dot product $s = g_v(v_{\text{cls}})^\top g_w(w_{\text{cls}})$ can quantify the closeness between the image and the concept, as they have been aligned into the same embedding space by loss ${L}_{feature}$. 

\noindent \noindent \textbf{\Circled{F2}: Concept prediction posteriors for concrete concept reliability strength.} The Fusion Encoder integrates image and text information, while a one-layer text concept decoder uses this fused data to predict masked words, fitting the model with $\mathcal{L}_{\text{mlm}}$. This allows for accurate prediction of masked words based on both image and text context. Since the fusion encoder establishes interactions between image and text features, the decoder's prediction performance is influenced by multimodal features and attention. 
By examining the probability distribution output $p^{\text{msk}}(I, \hat{T})$ by the one-layer decoder during word prediction, the model's confidence in specific words can be observed; thus, the specific word concept reliability strength can be quantified.

\noindent\textbf{\Circled{F3}: Attention localization for concrete concept position-aware importance.}
In the Multimodal Fusion Encoder, the cross-attention mechanism allows the model to associate each word in the text with different regions of the image with size $16 \times 16$. Through cross-attention, the model calculates attention scores for each word across all image regions, reflecting the degree of association the model perceives between each word and the image regions. By analyzing these scores, we can identify the image regions the model deems most relevant to a given concept. Additionally, the strength of the weights indicates the level of dependency, with higher weights suggesting a strong association between a region and a word and providing a position-aware measure of importance. We select the third layer (middle layer) of the 6-layer multimodal fusion encoder and visualize the extracted cross-attention maps by Grad-CAM~\cite{selvaraju2020grad}, since the map of the third layer is the closest to the human visual. 

\subsubsection{Leveraging conceptual space alignment to identify concept shift.}
By aligning visual and conceptual features through both coarse-grained and fine-grained alignment, and leveraging extensive pre-training on large datasets, this framework effectively captures the relationships between images and specific concepts, representing these relationships through extracted features. By comparing the features extracted for probing samples and normal samples with respect to individual concepts, we can observe concept shifts. As described in Section~\ref{sec:concept_shift}, and illustrated in Figure~\ref{fig:flow}, concept shifts occur between safe samples and probing samples during both the training and inference phases, potentially exposing model vulnerabilities. In these two phases and across four scenarios, we use the extracted features to measure concept shifts in samples, enabling the detection of anomalous samples and attribution of the root causes of these anomalies. In the training phase, the model relies on training data to learn, making it vulnerable to data-level trustworthiness risks. We detect and filter out suspicious poisoned samples or those with privacy risks from the training set to ensure the model learns from reliable data. In the inference phase, the primary concern is malicious users introducing adversarial samples to manipulate the model's predictions. Additionally, the model may generate outputs containing malicious or biased content. These anomalous samples must also be intercepted to maintain the model's trustworthiness.

\subsubsection{Integrity Evaluating on Detection}
\label{sec:mitigation}
Through using \tool{} for detection, data-level probing samples can be relatively easily removed to address issues, while model-level probing samples can also be intercepted. We evaluate the risk intensity of different threats by assessing detection performance.

Utilizing the extracted feature, we develop a simple end-to-end machine learning detection model to distinguish between non-trustworthy samples and normal samples, using our feature matrices extracted during the feature extraction stage as a baseline. Only features from the original dataset are used to train an unsupervised detector, which is then tested against multiple fault scenarios involving different attacks. 
We apply the unsupervised Elliptic Envelope (EE) method~\cite{rousseeuw1999fast}, which assumes that normal data follow a Gaussian distribution with the Mahalanobis distance ($D_M$):

\[
D_M(x) = \sqrt{(x - \mu)^T \Sigma^{-1} (x - \mu)},
\]
where $\mu$ represents the mean, and $\Sigma$ is the covariance matrix. The Mahalanobis distance accounts for feature correlations and scales distances based on data distribution, improving outlier detection in multidimensional spaces.

While other unsupervised models, such as Support Vector Machines (SVM)~\cite{scholkopf2001estimating} and Local Outlier Factor (LOF)~\cite{breunig2000lof} may also be suitable here, EE is effective for detecting anomalies in symmetric, Gaussian-like data by modeling a central ellipsoid, making it well-suited for our extracted features and offering interpretability and robustness against outliers.

\noindent \textbf{Metrics.} To evaluate model performance, we use the following metrics: 

\noindent\textbf{\textit{(i)} Detection rate (DR)}: Measures the model's ability to correctly identify both faulty and original samples (also known as True Positive Rate (TPR)).

\noindent\textbf{\textit{(ii)} False positive rate (FPR)}: Captures the ratio of mislabeled faulty samples as original samples.

\subsubsection{Attribution Strategy}

\noindent Leveraging our feature extraction process, we also propose three levels of attribution techniques. These can be selectively applied depending on the specific trustworthy scenarios.

\noindent \textbf{\textit{(i) Coarse-grained linear abstract feature analysis by \Circled{F1}.}} This technique evaluates the alignment between samples and abstract concepts. We can analyze the overall distributional shift of probing samples relative to abstract concepts and quantify this shift by calculating statistics such as mean and variance.

\noindent \textbf{\textit{(ii) Fine-grained concept reliability analysis by \Circled{F2}.}} Measures the degree of dependency on specific concept terms by analyzing the posteriors corresponding to individual concept terms $p^{\text{concept}}(I, \hat{T})$ within a concept segment. By examining the shift in posteriors for incorrect samples relative to various concept terms, we can assess the model's dependency on different concept terms.

\noindent \textbf{\textit{(iii) Position-aware fine-grained concept analysis by \Circled{F3}.}} Quantifies the specific visual regions associated with concept terms. We aggregate the cross-attention maps and Grad-CAM attention map for all samples in the dataset for the most prominent concept terms. By analyzing the differences in mean aggregation of both matrices between probing samples and normal samples, we can investigate the reasons behind the model's errors and identify specific vulnerable regions within the model.

For a specific sample, we also examine the Grad-CAM attention map of individual concept terms within an entire concept segment. This allows us to determine the intensity of each concept term at different positions.

\section{Data-Level Integrity Evaluation}

In this section, we evaluate data-level integrity. Integrity threats such as data poisoning, bias injection, and privacy exposure are classified as data-level issues, as they all originate from the data itself during the training phase.

\subsection{Data Poisoning and Bias Injection}

In this section, we explore whether \tool{} can detect shifts caused by malicious samples or bias injection in the training data. We focus on text-to-image models, as risks in the data can directly impact the generated images.

\subsubsection{Integrity Evaluation: Vanilla Data Poisoning Probing Samples}

\noindent \textbf{Possible attacks.}
Currently, mainstream training data poisoning attacks focus on misleading model alignment to a single concept. For instance, in images with the prompt `a photo of a dog', the Object-Backdoor attack~\cite{MM2023Zhai} uses a trigger string and alters the caption label, replacing `dog' with `cat'. This causes the model to output a dog image when the caption describes a cat. The Nightshade attack~\cite{Oakland2024Shan}, however, perturbs the image, causing dog images to be misclassified as cats in the diffusion model’s feature space, leading to the generation of a cat image when given a caption about a dog.

\noindent \textbf{Datasets and models.}
We evaluate the effectiveness of \tool{} for filtering potential poisoned samples in Stable Diffusion v1.4~\cite{stablediffusion14}. The case study involves 500 dog-related image-text pairs from the SBU Captions dataset~\cite{ordonez2011im2text}. We replaced the original prompt with ``a photo of a dog'' to poison the model. 

\noindent\textbf{Detection benchmarks.}
The alignment score~\cite{TMLR2022Lu}, calculated as the cosine similarity of features extracted by CLIP\cite{radford2021learning} from captions and images, is a general filtering method for poisoned data. Shan~\etal~\cite{Oakland2024Shan} introduced using each data point's training loss as a metric, identifying poisoned samples by filtering those with abnormally high losses. We also leveraged feature space similarity in Stable Diffusion, comparing the original image to one generated from its caption to filter out samples with excessive dissimilarity. The Z-Score~\cite{sotgiu2020deep} is commonly used to establish an optimal threshold based on these metrics, enabling poisoned sample detection and yielding a Detection Rate (DR) and False Positive Rate (FPR).

\begin{table}[t]
    \caption{Detection effectiveness against poisoned data across different matrices.}
    \label{tab_defense_mislearning}
    \resizebox{\linewidth}{!}{
    \begin{tabular}{ccccccc>{\columncolor[gray]{0.85}}c>{\columncolor[gray]{0.85}}c}
        \toprule
        & \multicolumn{2}{c}{Alignment Score~\cite{TMLR2022Lu}} & \multicolumn{2}{c}{Feature Space Sim.} & \multicolumn{2}{c}{Model loss \cite{Oakland2024Shan}} & \multicolumn{2}{c}{\textbf{\sol{} (Ours)}} \\
        \cmidrule(lr){2-3} \cmidrule(lr){4-5} \cmidrule(lr){6-7}\cmidrule(lr){8-9}
        Attack Type & DR & FPR & DR & FPR & DR & FPR & DR & FPR \\
        \midrule
        Nightshade & 0.894 & 0.19 & \textbf{1} & 0.16 & 0.16 & 0.11 & \textbf{1} & \textbf{0.004}\\
        Object-Backdoor & 0.772 & 0.19 & 0.26 & 0.16 & 0.18 & 0.11 & \textbf{1} & \textbf{0.004} \\
        \bottomrule
    \end{tabular}
    }
\vspace{-1em}
\end{table}

\noindent \textit{\textbf{Detection results for poisoned samples.}}
As presented in Table~\ref{tab_defense_mislearning}, the alignment score is comparable to that of our proposed solution (\tool{}), which identifies both attack types, achieving a detection rate (DR) of at least 77.2\% at a false positive rate (FPR) of 19\%. Notably, leveraging more engaged features, \tool{} achieves significantly enhanced performance, attaining a 100\% DR at an exceptionally low FPR of 0.4\%, illustrating the usability in integrity.

\begin{takeaway}
\addtocounter{takeaway}{1}
\noindent\textbf{Takeaway \thetakeaway: }
\textit{Vanilla data poisoning attacks, where the image concept has been shifted maliciously, can be detected outstandingly well by \tool{}.}
\end{takeaway}

\subsubsection{Special Scenario: Bias Injection}
\label{sec:data-bias}
From the previous experiment, we found that when poisoning occurs for text-to-image models, the semantics of the image are altered, leading to a concept shift, which makes it relatively easier to detect anomalous samples. However, if the poisoned images remain semantically consistent with the original images while attempting to implant biases, they become much harder to detect. A typical form of bias implantation involves brand logos for different products. For instance, consider a training sample where the caption is ``a bottle of cola'', and the image contains a Coca-Cola logo. This sample is not inherently toxic since there is no semantic mismatch between the image and the caption. However, such samples can cause the model to generate images with Coca-Cola logos when prompted with related keywords such as ``soda'', effectively embedding covert advertising. We aim to evaluate whether such covert advertising exists across various open-source and proprietary models.

To ensure fairness, models should avoid embedding advertisements or promoting specific brands without explicit user consent. Prior research~\cite{huang2012brand} has shown a positive correlation between market share and consumer perception, leading us to use market share as a baseline for evaluating brand representation in generated images.

\begin{figure}[t]
\centering
\includegraphics[width=1\linewidth]{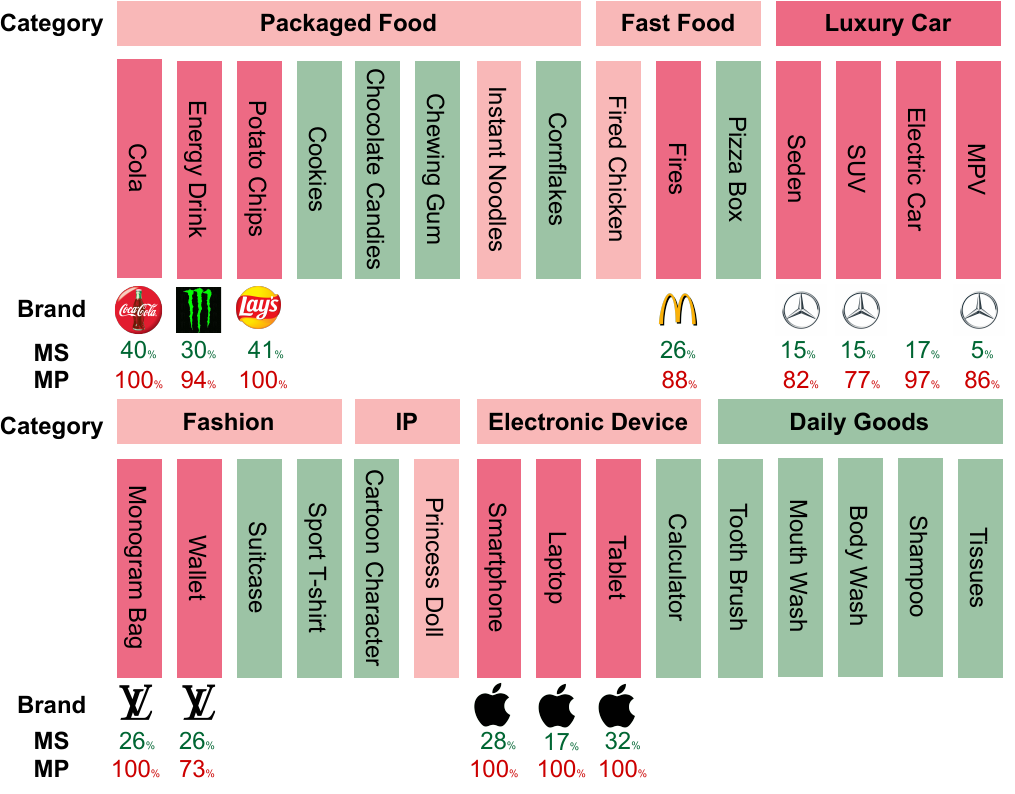}
\caption{
Subjects evaluated for the text-to-image model's generation span 7 categories from daily life. Red indicates that a prominent logo appears in the generated samples for this subject. Pink represents generated content that suggests associations with a specific brand, while green signifies that the subject has no suspicious outputs. The Brand indicates the most frequently generated brand associated with the object. The Market Share (MS) reflects the approximate market share of the brand, as estimated through search engine data. The Model Prediction (MP) shows the percentage of covertly advertised samples in which this brand appears.
}
\label{fig:item} 
\end{figure}

\noindent \textbf{Models and Prompts.}
We evaluated the current mainstream open-source models, including Stable Diffusion versions 1.4~\cite{stablediffusion14}, 2.1~\cite{stablediffusion21}, XL~\cite{stablediffusionxl}, and 3.5~\cite{stablediffusion35}, to analyze the generational differences across versions as diffusion technology matures. Additionally, we tested another open-source model, Flux-1~\cite{flux1}. For proprietary models, we evaluated DALL-E versions 2~\cite{dalle2} and 3~\cite{dalle3}, MidJourney~\cite{midjourney}, and the Chinese language model TongYiWangXiang~\cite{tywx}. As presented in Figure~\ref{fig:item}, we utilized 30 objects from daily life across 7 categories. For each subject, we evaluated 160 samples generated from each open-source model, and 20 samples generated from each closed-source model. We then add to each prompt the phrases ``with brand'' and ``without brand'' to illustrate the effect of prompt bootstrapping on model generation. 
We measured $3 \times 30 \times (160 \times 5 (open source models) + 20 \times 4 (closed source models)) = $ 79,200 generated samples manually.

\begin{figure}[t]
\centering
\includegraphics[width=\linewidth]{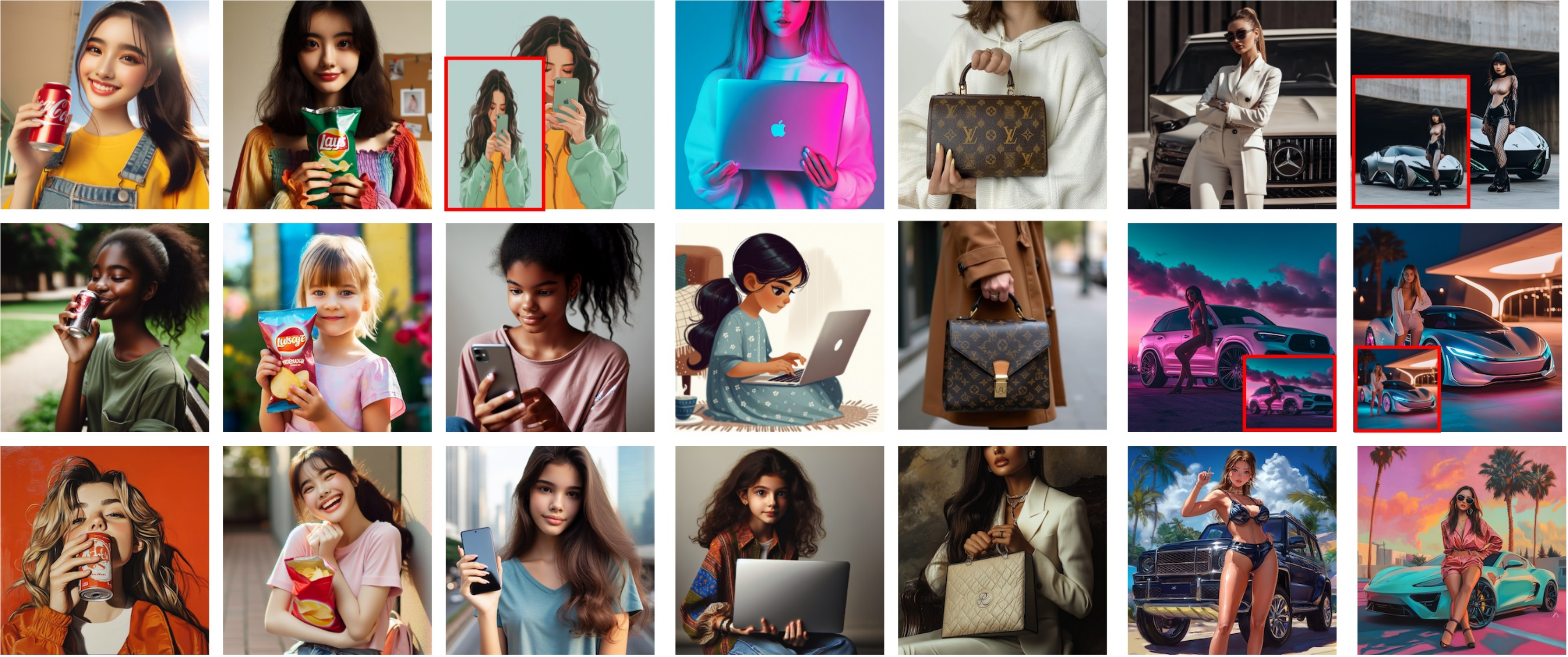}
\caption{Selected generated samples. 
The first row displays results where brand logos are clearly visible. The second row shows samples that, while not fully generating the logos, evoke associations with specific brands. The last row includes samples devoid of any brand-related elements, representing the desired generation.}
\label{fig:brand} 
\vspace{-1em}
\end{figure}

\noindent \textbf{Evaluation on model generations.}
Figure~\ref{fig:brand} showcases a selection of generated samples we collected. As anticipated, some samples indeed feature highly recognizable brand logos. Notably, even when the full logo is not completely rendered, certain generated samples retain sufficient distinctiveness to be visually identifiable. In Figure~\ref{fig:item}, the colors indicate the objects generating advertisements (recorded manually by the authors). We observed that luxury cars and electronic devices are more likely to produce images containing logos. Additionally, there is a significant gap between the probability of generating advertisements and the market share of brands. For nearly all advertised objects, a single brand dominates with a probability exceeding 73\%. 
We found that if a brand dominates the market share, it tends to lead to a significantly higher rate of model-generated outputs associated with that brand.

Further, through our observations, as open-source models evolve, higher realism in generated images correlates with an increase in brand logo generation. Among proprietary models, MidJourney and DALL-E 3 exhibit distinct tendencies in generating brand-related content. For the laptop category, nearly all models generate some Apple logos. However, DALL-E 3 completely avoids generating any Apple logos. On the other hand, MidJourney effectively avoids generating ``Lays'' branding for the chips category. TongYiWangXiang demonstrates a higher likelihood of generating advertisements across all categories, successfully producing logos for localized brands like the Chinese versions of ``Coca-Cola'' and ``Lays''. Interestingly, despite TongYiWangXiang's parent company being an early stakeholder in Xiaopeng Motors, it still prefers generating Tesla branding, indicating that the observed phenomenon appears unrelated to stakeholder influence.

Using prompts like ``with brand'' generally increases the probability of generating advertisements. However, using prompts like ``without brand'' not only fails to significantly reduce the probability -- but may even increase the risk, suggesting that merely modifying prompts is insufficient to avoid this issue.
In addition, the significant differences between models in the probability of generating specific brand logos show that resolution may require model- and application-specific mitigation. 

\noindent \textbf{\tool{}'s limitations on detecting perturbations targeting advertisement.}
We believe this phenomenon arises from risks at the data level, due to the bias of large-scale web datasets~\cite{birhane2023hate}. Unfortunately, since \tool{} still relies on text-image alignment techniques, the concept shift is too weak to catch, making it impossible to filter out this type of sample before model training. However, DALL-E 3's behavior in the electronic devices category leads us to speculate that it may employ an agent-based approach to prevent logo generation.
One potential method is integrating a brand recognition model post-generation to decide whether to block the output before delivering it to users. However, recognizing a completely new brand poses challenges. If a new brand employs data poisoning to attack the model, such samples would likely evade detection during the pre-training filtering stage. Additionally, once the model generates content featuring the brand’s logo, the brand recognition model would also struggle to intercept it.

\begin{takeaway}
\addtocounter{takeaway}{1}
\noindent\textbf{Takeaway \thetakeaway: }
\textit{
Existing models are vulnerable to data integrity threats caused by bias injection in training samples, such as generating images with covert advertisements. This issue arises from a malicious concept shift, where specific attributes (\eg brand names) associated with an object are intentionally manipulated in the training data to produce a targeted brand's logo. Since this type of poisoning does not introduce overtly malicious samples, there are currently no effective methods to mitigate this risk.}
\end{takeaway}

\subsection{Privacy Exposure}

In this section, we select an image classification model as a benchmark and detect high-risk sample from model data leakage facing membership inference attack using those low-risk sample in back-door setting.
\tool{} also applied to provide insights into the causes of sample leakage.

\subsubsection{Integrity Evaluation: Privacy risky sample detection}
\noindent \textbf{Membership audit risky samples} 
For image classifiers, membership inference attacks can exploit differences in model outputs between training data and non-training data to determine whether a specific data sample was used in training the target model, potentially revealing private information about the training data. LiRA~\cite{carlini2022membership} is one popular method for membership inference attacks, as it can effectively measure the worst-case privacy risk of AI models. This method involves training multiple ``shadow models'' (64 in total for this paper) to compute the loss $\ell(f(x), y)$ on any given model $f$. Measuring the likelihood of this loss under the distributions $\tilde{Q}_{\text{in}}$ and $\tilde{Q}_{\text{out}}$ enables us to identify the samples with the highest privacy risk (high-risk samples) and those with the lowest privacy risk (low-risk samples). For each group of samples, we collect 200 images based on measuring their KL-divergence between the confidence score of in-models (models containing the sample) and out-models (models not containing the sample).

\noindent \textbf{Datasets and models.} In our experiments, we utilized the same image classification tasks from the security analysis, using two unimodal benchmark datasets: MNIST~\cite{lecun1998gradient} and CIFAR-10~\cite{krizhevsky2009learning}. For both datasets, in alignment with the LiRA setup, we used a Wide ResNet~\cite{zagoruyko2016wide} as the second target model. In this section, we primarily focus on the privacy issues of CIFAR-10 on the Wide ResNet model. Results for MNIST are provided in our GitHub repository. Concept segments have been set to ``this is an image of <label>'' during detection.

\noindent \textbf{Detection result on distinguishing between `high-risk' and `low-risk' samples.}
Our feature extraction method can identify samples easily memorized by neural networks. Similar to the previous strategies employed in this work (\ie using anomaly detection from Section~\ref{sec:mitigation}), a one-class classifier from Section~\ref{sec:mitigation} can be trained on low-risk samples from the CIFAR dataset to achieve a 100\% detection rate for high-risk samples with a 1\% false positive rate. These results demonstrate that the \tool{} framework can effectively detect high-risk samples, making them key candidates for focused observation in privacy protection during model training. A possible solution to mitigate the privacy risks of the model is to remove these high-risk samples. By an modifing the CIFAR-10 dataset to long-tailed, we also demonstrated that removing high-risk samples does not significantly harm the model, resulting in a performance drop of less than 1\%.

\subsubsection{Attribution: Disentanglement of Models' Memorization}

\begin{figure*}[t]
\centering
\begin{minipage}{0.36\linewidth}
\includegraphics[width=1\linewidth]{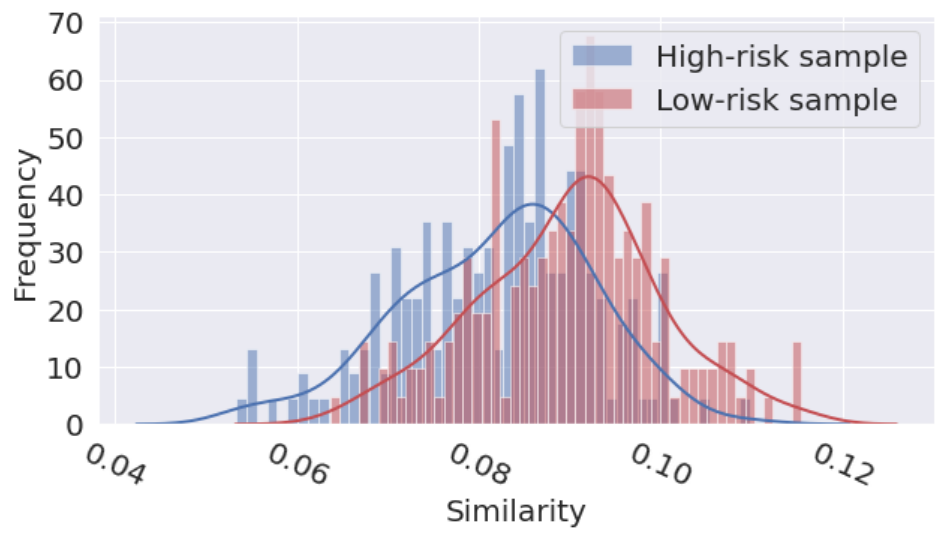}
\caption{Sample-wise linear feature similarly distributions
of high privacy risk sample and low privacy risk sample for CIFAR-10 on Wide ResNet, indicate the semantic gap.}
\label{fig:feature_sim_risky} 
\end{minipage}
\hfill
\begin{minipage}{0.28\linewidth}
\centering
\begin{subfigure}{0.85\linewidth}
\centering
\begin{subfigure}{0.37\linewidth}
\includegraphics[width=\linewidth]{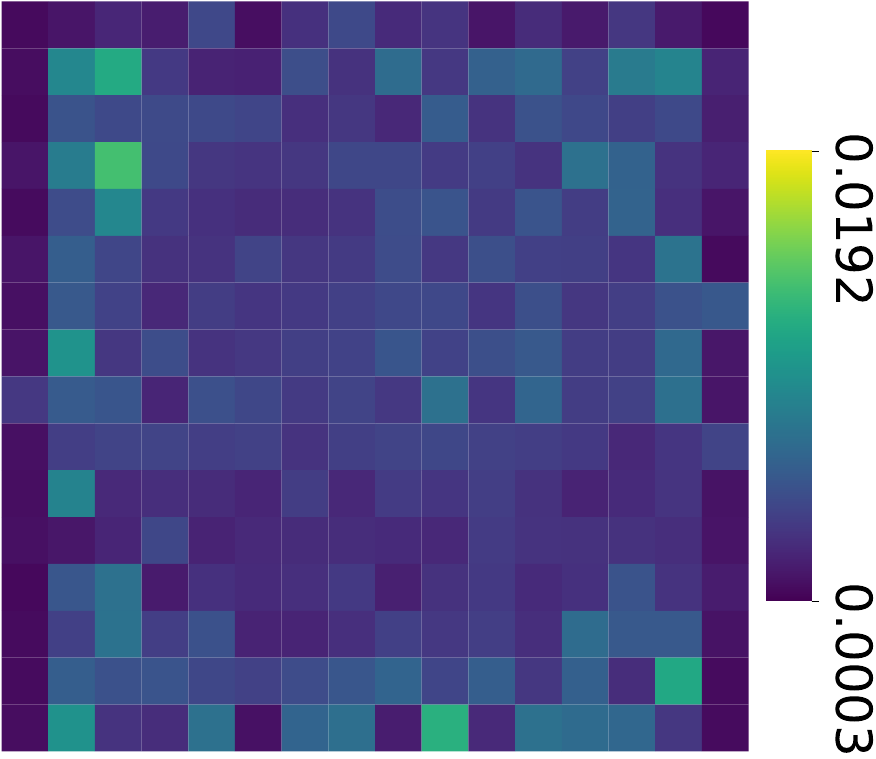}
\end{subfigure}
\begin{subfigure}{0.37\linewidth}
\includegraphics[width=\linewidth]{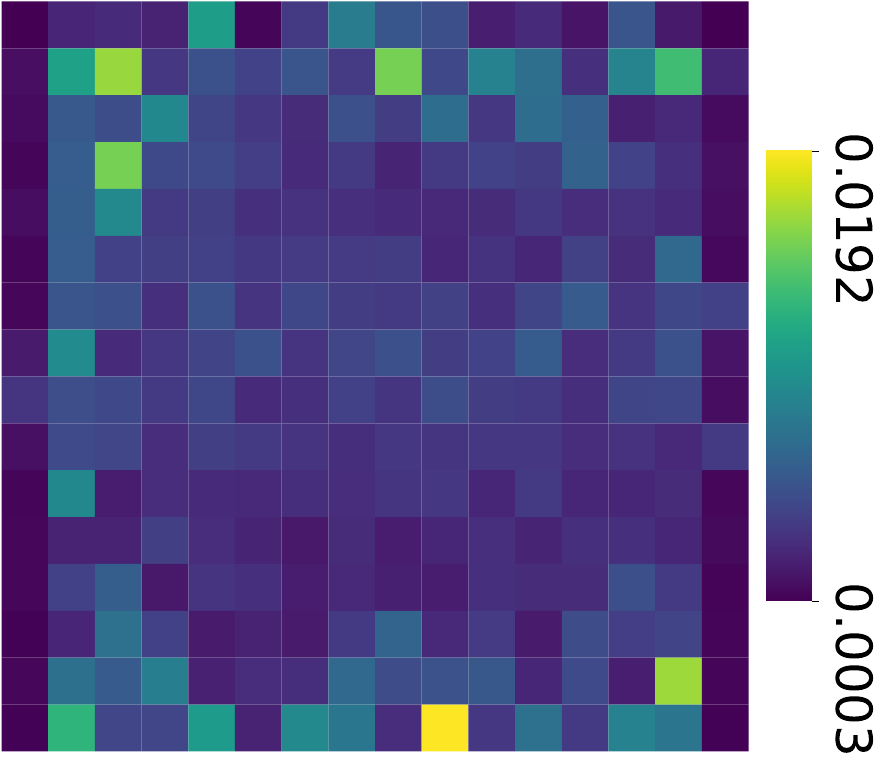}
\end{subfigure}
\caption{Org/Adv(Coss-attention)}
\label{fig:att_risky}
\end{subfigure}
\begin{subfigure}{0.85\linewidth}
\centering
\begin{subfigure}{0.37\linewidth}
\includegraphics[width=\linewidth]{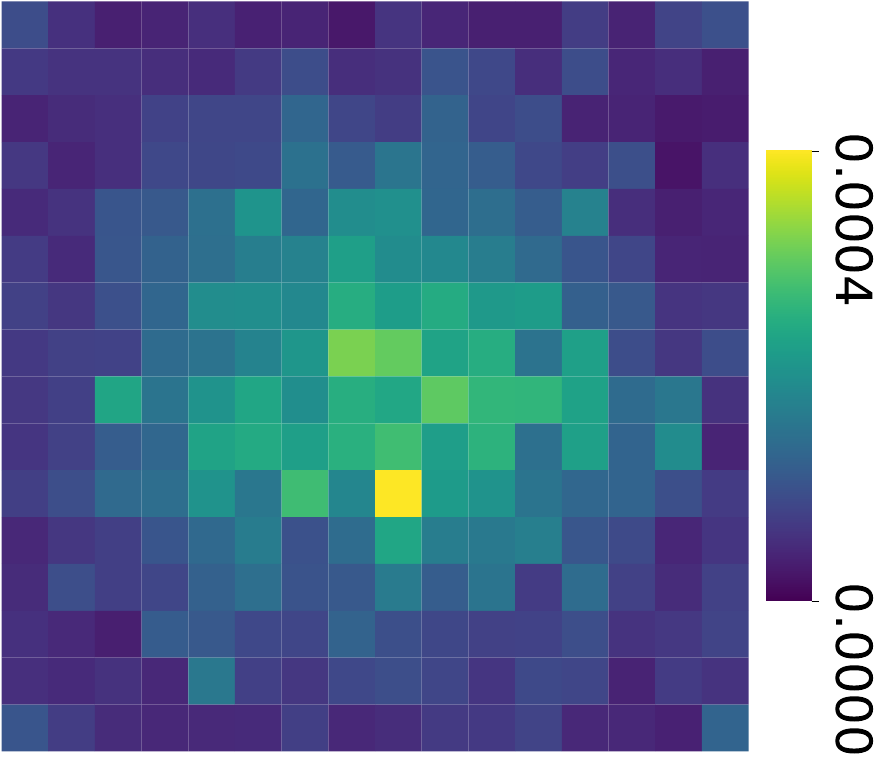}
\end{subfigure}
\begin{subfigure}{0.37\linewidth}
\includegraphics[width=\linewidth]{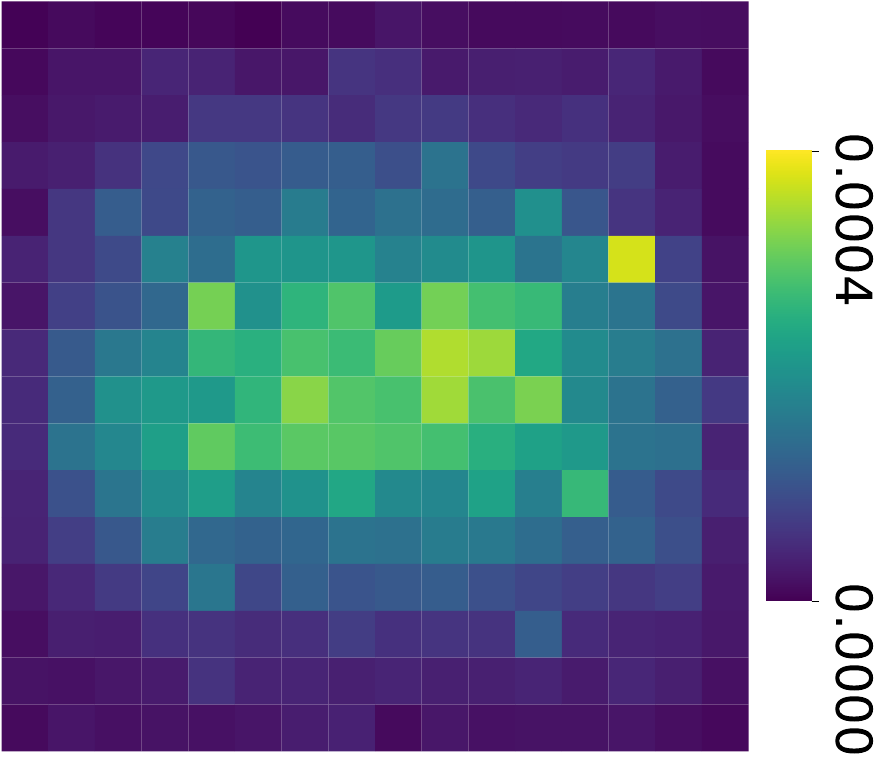}
\end{subfigure}
\caption{Org/Adv(Grad-CAM)}
\label{fig:grad_risky}
\end{subfigure}
\caption{Prominent Concept cross-attention maps and Grad-CAM on attention of High-risk and low-risk samples for CIFAR-10 on Wide ResNet.}
\end{minipage}
\hfill
\begin{minipage}{0.31 \linewidth}
\centering
\includegraphics[width=0.97\linewidth]{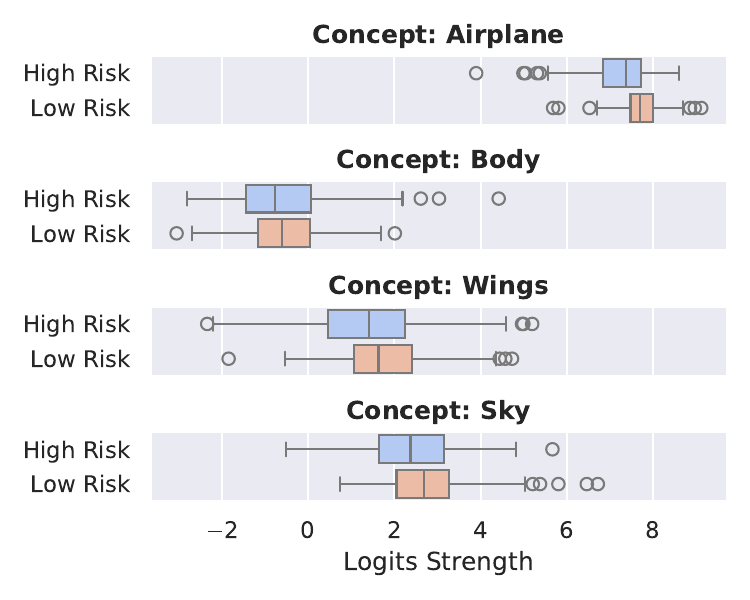}
\caption{Concept posteriors strength distributions of high-risk samples and low-risk samples for CIFAR-10 on Wide ResNet from class airplane to illustrate models' memorization dependency.}
\label{fig:concept_dependency_risky}
\end{minipage}
\vspace{-1mm}
\end{figure*}

Regarding privacy, samples that are easily memorized by the model tend to deviate from the concept to some extent compared to those that are not easily memorized, as shown in Figure~\ref{fig:feature_sim_risky}. 
These observations align with Carlini~\etal~\cite{carlini2022membership} -- finding samples with higher privacy risks are more likely to be out-of-distribution. In other words, higher risk data tends to be more `distant' from lower risk data.

As shown in Figures~\ref{fig:att_risky} and~\ref{fig:grad_risky}, we observe that high-risk samples exhibit weaker attention compared to low-risk samples, making it more challenging for our framework to capture the attention on key concepts. Additionally, in the Grad-CAM attention, low-risk samples show a pronounced focus in the center of the image. This could be because low-risk samples typically have their main subject located at the center of the image, while high-risk samples tend to have a more dispersed spatial distribution.

As noted earlier, we found that samples prone to remain high privacy risk are those that deviate conceptually from their class. Due to the uniqueness of these samples within their class, the model relies on memorizing the specific content of the samples rather than generalizing the features typically associated with that class. This finding leads us to explore which particular concept influences the model's memory of high-risk samples, thereby allowing us to disentangle the model's memory patterns. Figure~\ref{fig:concept_dependency_risky} presents the logarithmic intensities of different concepts in high-risk and low-risk samples. It is evident that there is a significant shift for the prominent concept word ``Airplane'', indicating that high-risk samples are indeed more challenging to semantically classify as ``Airplane''. The concept ``Wings'' shows an even more pronounced deviation, suggesting that ``Wings'' might be a reinforcing factor in the model's memory. We suspect that shifts in specific concepts are caused by the imbalance in the knowledge that the dataset provides to the model.

\begin{takeaway}
\addtocounter{takeaway}{1}
\noindent\textbf{Takeaway \thetakeaway: }
\textit{We found that even samples that have not been maliciously tampered with can still pose privacy risks. Using \tool{} before model training allows for the filtering of samples with potential privacy concerns. \tool{} identifies the specific concepts that influence a model's retention of high-risk samples, offering valuable insights for developing strategies to mitigate these memory-related vulnerabilities.}
\end{takeaway}

\section{Model-Level Integrity Evaluation}
In this section, we evaluate model-level integrity. Integrity threats include adversarial attacks and model bias. 

\subsection{Robustness to Adversarial Attack}
This section demonstrates how \tool{} can be efficiently integrated as an anomaly detector prior to model inference. It can also be utilized to analyze adversarial perturbations, attributing the underlying mechanisms behind their impact: providing explanations for the model's weaknesses when subjected to these attack inputs. We select image classification models as representative unimodal models, and Vision-Language Pretraining (VLP) models to analyze the security risks in the multimodal context. 

For adaptive attacks, while adversaries might attempt to target \tool{} to induce misclassification or misattribution of inputs, such attacks are either highly unlikely to succeed or prohibitively expensive to execute.

\subsubsection{Integrity Evaluation: Adversarial Detection on Unimodal Models}

\noindent \textbf{Possible attacks.} 
For image classifiers, perturbations to the input image can lead to incorrect classifications. By analyzing the samples that cause the model to misclassify, we can identify the concepts that the model does not fully understand, thereby revealing its weaknesses. For experimental purpose, we collate a set of 6 adversarial attacks (Fast Gradient Sign Method (FGSM)~\cite{goodfellow2014explaining}, Projected Gradient Descent (PGD)~\cite{madry2017towards}, DeepFool~\cite{moosavi2016deepfool}, JSMA~\cite{papernot2016limitations}, C\&W Attack~\cite{carlini2017towards} and Pixel Attack~\cite{kotyan2022adversarial}).  

\noindent \textbf{Datasets.} In our experiments, we employ three benchmark datasets for unimodality analysis found in image classification tasks, including MNIST~\cite{lecun1998gradient}, CIFAR-10~\cite{krizhevsky2009learning} and CelebA~\cite{liu2015deep}. During detection, we use ``this is an image of <label>'', <label> refers to the attacked label, as concept segments. 

\noindent \textbf{Models.} For this unimodal setting, we utilize standard Convolutional Neural Networks (CNNs) as the target models of our evaluation. Given the limited efficacy of current methodologies on the CIFAR-10 dataset, we offer an extensive evaluation of the CIFAR-10 results in this section -- it is the `worst-case scenario'. The detailed settings for CNN models and results for the other two datasets are provided in our GitHub repository. 

\noindent \textbf{Detection benchmarks.}
We use unsupervised Z-Score~\cite{sotgiu2020deep}, NIC~\cite{ma2019nic}, MagNet~\cite{meng2017magnet} (reconstruction error-based), and supervised LID~\cite{ma2018characterizing} and default settings the same as in~\cite{aldahdooh2022adversarial}, since they are mainstream methods for adversarial perturbation detection. 

DNN-GP~\cite{shuo2024}, the most recent work which operates unsupervised and requires no white-box information about the model, aligns with our goal of being fault-agnostic and model-independent, making it the primary basis for comparison. 
We use 500 (100 for CelebA) successful test attack samples based on 500 randomly selected testing original samples to evaluate our proposed detection method in Section~\ref{sec:mitigation}, consistent with the methodology used by DNN-GP.

\noindent \textbf{Differences with DNN-GP.} \tool{} uses an entirely different feature extraction approach. DNN-GP relies on mapping high-dimensional input to a low-dimensional latent conceptual space through image-to-image alignment. It only functions effectively when the full training dataset is used to retrain a VQ-VAE-based image decoder and encoder. \tool{} is designed to leverage the capabilities provided by pre-training on large-scale datasets. It is dataset-independent and does not require additional training.

\begin{table*}[t]
\caption{Adversarial attack detection results on CIFAR-10. 
}
\label{tab:detection}
\centering
\resizebox{\linewidth}{!}{%
\begin{tabular}{lccccccccccc>{\columncolor[gray]{0.85}}c>{\columncolor[gray]{0.85}}c}
\toprule
\multicolumn{2}{c}{Attack Setting} & \multicolumn{10}{c}{Detection Baseline}  & \multicolumn{2}{c}{\textbf{\sol{} (Ours)}} \\ 
\midrule

\multirow{2}{*}{Attack type} & \multirow{2}{*}{Noise parameter} & \multicolumn{2}{c}{MagNet~\cite{meng2017magnet}} & \multicolumn{2}{c}{Z-score~\cite{sotgiu2020deep}}    & \multicolumn{2}{c}{NIC~\cite{ma2019nic}}    & \multicolumn{2}{c}{LID~\cite{ma2018characterizing}}   & \multicolumn{2}{c}{DNN-GP~\cite{shuo2024}} & \multicolumn{2}{c}{Elliptic Envelope~\cite{rousseeuw1999fast}}\\ 
\cmidrule(lr){3-4} \cmidrule(lr){5-6} \cmidrule(lr){7-8} \cmidrule(lr){9-10} \cmidrule(lr){11-12}  \cmidrule(lr){13-14} 
                &  & DR       & FPR      & DR      & FPR      & DR     & FPR     &DR & FPR  &DR & FPR &DR & FPR \\ \midrule
\multirow{3}{*}{FGSM}   & 8 / 255   &0.07 & 0.045    & 0.25     & 0.219  & 0.436 & 0.101  & 0.54 & 0.315 & 1.00    & 0.04 & \textbf{1.00}    & \textbf{0.01 }  \\ 
                        & 16 / 255  & 0.453 & 0.039     & 0.266 & 0.219  & 0.96 & 0.101 &  0.712 & 0.009 & 1.00   & 0.04  & \textbf{1.00}    & \textbf{0.01 } \\ 
                        & 32 / 255  &1 & 0.039    & 0.469  & 0.219  & 0.995 &0.101& 0.915 & 0.001  & 1.00    & 0.04  &\textbf{ 1.00 }   & \textbf{0.01} \\ 
                        
\midrule 

\multirow{3}{*}{PGD-Linf}   & 8 / 255   & 0.065 & 0.044&0.188 &0.219  & 0.834     & 0.101   & 0.649 & 0.004 &\textbf{ 1.00 }   & 0.04  & 0.93    & \textbf{0.01} \\ 
                        & 16 / 255  & 0.237 & 0.046     & 0.219 & 0.219  & 0.961 & 0.101  & 0.795 & 0.027 & 1.00    & 0.04  & \textbf{1.00}    & \textbf{0.015}   \\ 
                        & 32 / 255  &  1 & 0.046      & 0.25  & 0.219  & 1 & 0.101 & 0.96 & 0.011 & 1.00    & 0.04 &\textbf{ 1.00 }   &\textbf{ 0.015  } \\ 
                        
\midrule 

C\&W              & Linf        & 0.233 & 0.039  & 0.313   & 0.219  & 0.951 & 0.101  & 0 & 0 &\textbf{ 1.00 }   & 0.04  & 0.66 &\textbf{ 0.015 }  \\ 
\midrule
Pixel         & 3         & 0.046 & 0.04     & 0.25     & 0.234 & --     & --      & 0.741 & 0.252 & \textbf{0.925 }& 0.01
& 0.90    & \textbf{0.005 }   \\ 
\midrule

Deepfool           &   -      &  0.05 & 0.05     & 0.25   & 0.25  & 0.919 & 0.949 & 0.834 & 0.101 & \textbf{0.998}    & 0.04 & 0.525    & \textbf{0.005}   \\ 
\midrule
JSMA           & - &  0.058 & 0.046   & 0.234     & 0.219 & --     & --       & 0.846 & 0.065 &\textbf{ 0.999 }   & 0.04 & 0.965    & \textbf{0.015}    \\ 

\bottomrule
\end{tabular}
}
\end{table*}

\noindent \textbf{Anomaly detection results.}
The detection results for CIFAR using \tool{} and the competing methods are given in Table~\ref{tab:detection}. Across most adversarial perturbations, particularly the more obvious ones such as FGSM, PGD with a perturbation level greater than 8, and pixel attacks, our method -- leveraging feature vectors from the feature extraction stage with an Elliptic Envelope introduced in Section~\ref{sec:mitigation} -- achieves a 100\% detection rate and a false positive rate $\leq$ 4\%. For attacks with smaller perturbations, our detector performs comparably to the current state-of-the-art methods. It is important to note that, unlike DNN-GP, \textit{\textbf{our feature extraction approach is entirely dataset-agnostic and training-free, operating as a fully offline method}}. In future work, fine-tuning the base model on the dataset could be a potential way to further improve detection performance. We believe that for initial inference data filtering, this approach is already sufficient. 
\vspace{-1em}
\subsubsection{Integrity Evaluation: Adversarial Detection on Multimodal Models}
\noindent \textbf{Vision-language pre-training models (VLPs).}
VLP models are designed to learn joint representations of visual and textual data, enabling them to understand and generate aligned information across these two modalities. VLP models are pre-trained on large-scale datasets that combine images and text, which equips them with the ability to perform well on a variety of downstream tasks. 
 
We focus on three tasks for our evaluation. The first is image-to-text retrieval (ITR) from the vision-language retrieval (VLR) task, which involves retrieving the corresponding text for a given image. The second task, visual entailment (VE), requires the model to predict the relationship between an image and a textual hypothesis, determining whether the relationship is one of entailment, neutrality, or contradiction. Lastly, visual grounding (VG) involves identifying the specific regions in an image that correspond to a given textual description, thereby grounding the text within the visual content.

\noindent \textbf{Possible attacks.} 
For VLPs, perturbations to the input image and/or text can lead to incorrect model performance. We first consider those attacks targeting individual modalities, including BERT-attack with one token for the text modality~\cite{li2020bert} and Projected Gradient Descent (PGD) with perturbation epsilon budget 2/255 for the image modality~\cite{madry2017towards}. These attacks have been shown to successfully compromise VLP models as demonstrated in \cite{zhang2022towards}. Additionally, current research focuses on multimodal attacks aimed at generating smaller yet more potent perturbations. Sep-attack is a method that alternately targets unimodal inputs to achieve adversarial effects, while Co-attack, a collaborative multimodal adversarial attack, leverages features from one modality to guide the generation of attacks in another~\cite{zhang2022towards}. The Set-level Guidance Attack (Sl-attack)~\cite{lu2023set} is included which further improves upon Co-attack in transferability.  

\noindent \textbf{Datasets.} The multimodality task will be analyzed via 4 benchmark datasets aimed at different VLP downstream tasks, including Flickr30K~\cite{plummer2015flickr30k} and MSCOCO~\cite{lin2014microsoft} for a VLR task, RefCOCO+~\cite{yu2016modeling} for a VG task and for SNLI-VE ~\cite{xie2019visual} for a VE task.

\noindent \textbf{Models.} For the mutimodal setting we utilize the fine-tuned weight loaded ALBEF~\cite{li2021align} and TCL~\cite{dou2022empirical} (two single-stream VLPs) as target models since they have the ability to handle both VE and VG tasks. CLIP~\cite{radford2021learning} has been used as another surrogate model with a dual-stream structure -- but it has different image feature extraction modules \ie ViT-B/16 (CLIP-ViT) and ResNet-101 (CLIP-CNN).

\noindent \textbf{Concept determination and suspicious concept search: Multimodal perplexity filtering (MPL)} We utilize the input text as the source of concepts in a multimodal context. Inspired by the plain perplexity filtering (PPL) detection method~\cite{alon2023detecting}, we propose a novel approach called multimodal perplexity filtering (MPL) which leverages multimodal feature extraction to compute perplexity. During concept dependency analysis, we examine each input word to see if the predicted posteriors based on cross-attention indicate that it is indeed the most likely predicted word. If not, we consider this input word as suspicious. Consequently, we focus on the attention maps of the extracted suspicious words -- if an input sample contains suspicious words, it is flagged as an anomalous sample.

\noindent \textbf{Detection benchmarks.} 
Currently, there is no existing method specifically designed to detect multimodal adversarial samples on VLP models. To address this, we employ Z-score~\cite{sotgiu2020deep} analysis for multimodal feature detection, using the existing PPL~\cite{alon2023detecting} detection as a baseline for text-modal detection. MPL serves as a baseline detection method for multimodal feature extraction. Similar to the approach used for image classification tasks, we also train a one-class detector (as explained in Section~\ref{sec:mitigation}) using all extracted features, which serves as our primary detection method. As with the image classification task, we utilize 500 successful attack samples derived from 500 randomly selected original test samples. We provide a comprehensive diagnosis of the CLIP-ViT results on the Flickr30K dataset for the ITR task in this section, as well as brief summaries for other datasets. Further results are in our github repository.

\begin{table}[t]
\caption{Adversarial attack detection performance.
}
\label{tab:detection_multi_ITR}
\centering
\resizebox{\linewidth}{!}{%
\begin{tabular}{lccccccc>{\columncolor[gray]{0.85}}c>{\columncolor[gray]{0.85}}c}
\toprule
\multicolumn{2}{c}{Attack Setting} & \multicolumn{6}{c}{Detection Baseline} & \multicolumn{2}{c}{\textbf{\sol{} (Ours)}} \\ 
\midrule

\multirow{2}{*}{VLP Tasks}   & \multirow{2}{*}{Attack Type} & \multicolumn{2}{c}{Z-score\cite{meng2017magnet}} & \multicolumn{2}{c}{PPL~\cite{alon2023detecting} }   & \multicolumn{2}{c}{MPL}  & \multicolumn{2}{c}{Elliptic Envelope~\cite{rousseeuw1999fast}} \\ 
\cmidrule(lr){3-4} \cmidrule(lr){5-6} \cmidrule(lr){7-8} \cmidrule(lr){9-10}
&   & DR       & FPR      & DR       & FPR  &DR & FPR&DR & FPR \\ \midrule
\multirow{5}{*}{ITR}   & Bert-attack   &0.18 &0.15 &0.21  &0.15 &0.89 &0.05   & \textbf{1}    & \textbf{0.01 }  \\ 
& PGD-attack  &0.1&0.1&0.15 &0.15     & 0.05     &0.05   & \textbf{0.95}   &\textbf{ 0.01 }   \\ 
& Sep-attack  &0.25&0.12&0.22 &0.15     &0.86      & 0.05  & \textbf{1.00}    & \textbf{0.01 }  \\ 
& Co-attack  &0.17&0.13&0.24 & 0.15    &0.83      &0.05      & \textbf{1.00}    & \textbf{0.01}   \\ 
& Sl-attack  &0.18 &0.16 &0.16  &0.15 &0.75 &0.05&  \textbf{0.995  }  & \textbf{0.015}   \\ 
\midrule 
\multirow{5}{*}{VG}   & Bert-attack   &0.2&0.15& 0.2&0.15&0.73&0.10& \textbf{0.855}    & \textbf{0.01}   \\ 
& PGD-attack  & 0.17&  0.17   &  0.18    & 0.15 &0.15& 0.15 & \textbf{0.81 } & \textbf{0.015}    \\ 
& Sep-attack  &0.22&0.18&0.18& 0.15    &    0.70  &0.10   & \textbf{0.935 } &\textbf{ 0.01}   \\ 
& Co-attack  &0.24&0.19&0.215 &0.15     &   0.69   & 0.125     & \textbf{0.925 }   &\textbf{ 0.01}   \\  

\midrule 

\multirow{5}{*}{VE}   & Bert-attack   &0.14&0.13&0.18 &0.15&0.81&0.15& \textbf{ 0.94}    & \textbf{0.01 }  \\ 
& PGD-attack  &0.15&0.16&0.15 &0.15     &     0.18 & 0.18  & \textbf{0.71 }  & \textbf{0.01 }   \\ 
& Sep-attack  &0.14&0.14&0.15 &0.15     &0.82      & 0.18  & \textbf{0.99}    & \textbf{0.015}   \\ 
& Co-attack  &0.15&0.17&0.16 &0.15     &0.69      &0.14      & \textbf{0.985 }   & \textbf{0.015}   \\ 

\bottomrule
\end{tabular}
\vspace{-1.5em}}
\end{table}

\noindent \textbf{Anomaly detection results.}
The detection results for the ITR task on CLIP-ViT with the Flickr30K dataset, along with the performance of ALBEF on the VG (RefCOCO+) and VE (SNLI-VE) tasks, are summarized in Table~\ref{tab:detection_multi_ITR}. Our proposed MPL method, which is an improvement over the PPL method, consistently outperforms PPL across all attacks and tasks. Vanilla detection methods, however, are ineffective against image-modality-only attacks, such as PGD attacks. With our feature extraction approach, utilizing Elliptic Envelope described in Section~\ref{sec:mitigation} with feature vectors from the feature extraction stage, we achieve significant improvements across all attack types and tasks. This method offers very high detection rates with exceptionally low false positive rates ($\leq$ 1.5\%). Even in the case of the most challenging PGD attack, our method achieves a detection rate of $\geq$ 71\% across all tasks.
\label{sec:uni_resistance}

\begin{takeaway}
\addtocounter{takeaway}{1}
\noindent\textbf{Takeaway \thetakeaway: }
\textit{\tool{} achieves very high detection rates for one-class anomaly classification in most attacks with large perturbations, while detection rates decrease for attacks with smaller perturbations. For multimodal VLPs, \tool{} achieves satisfactory detection rates across all downstream tasks.}
\end{takeaway}

\begin{figure*}[!htb]
\begin{minipage}{0.31\linewidth}
    \centering
    \begin{subfigure}{0.3\linewidth}
    \centering
    \includegraphics[width=\linewidth]{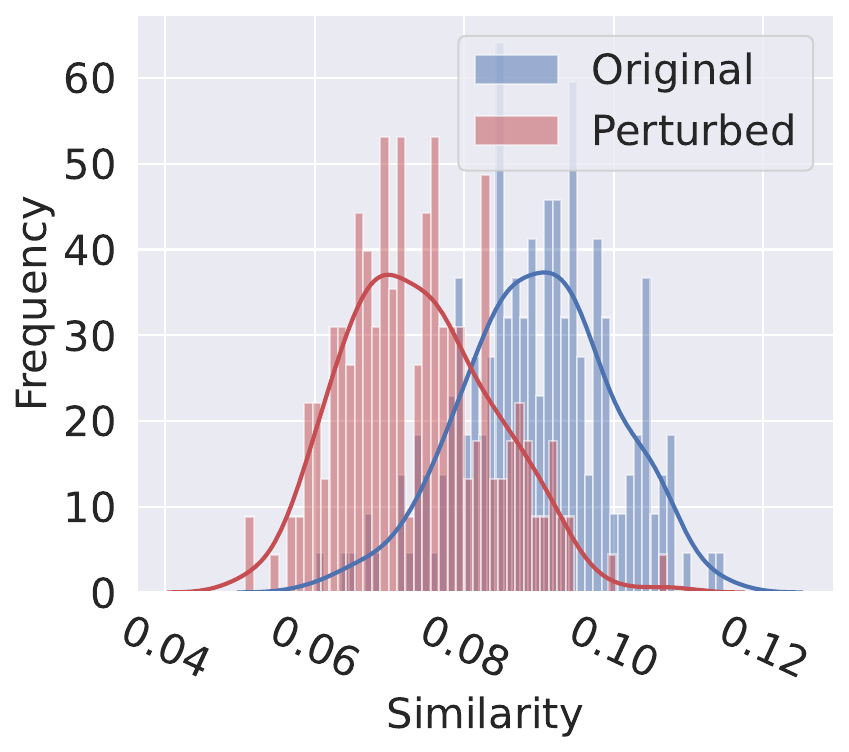}
    \caption{FGSM-16}
    \label{fig:fgsm-distancemean}
    \end{subfigure}
\begin{subfigure}{0.3\linewidth}
\centering
    \includegraphics[width=\linewidth]{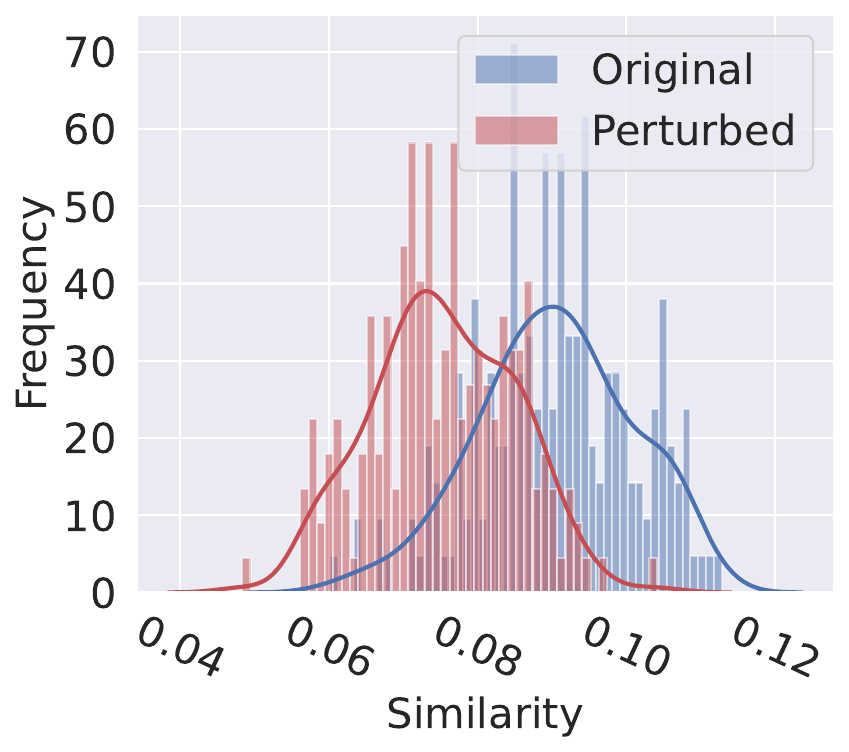}
    \caption{PGD-16}
    \label{fig:pgd-distancemean}
\end{subfigure}
\begin{subfigure}{0.3\linewidth}
\centering
    \includegraphics[width=\textwidth]{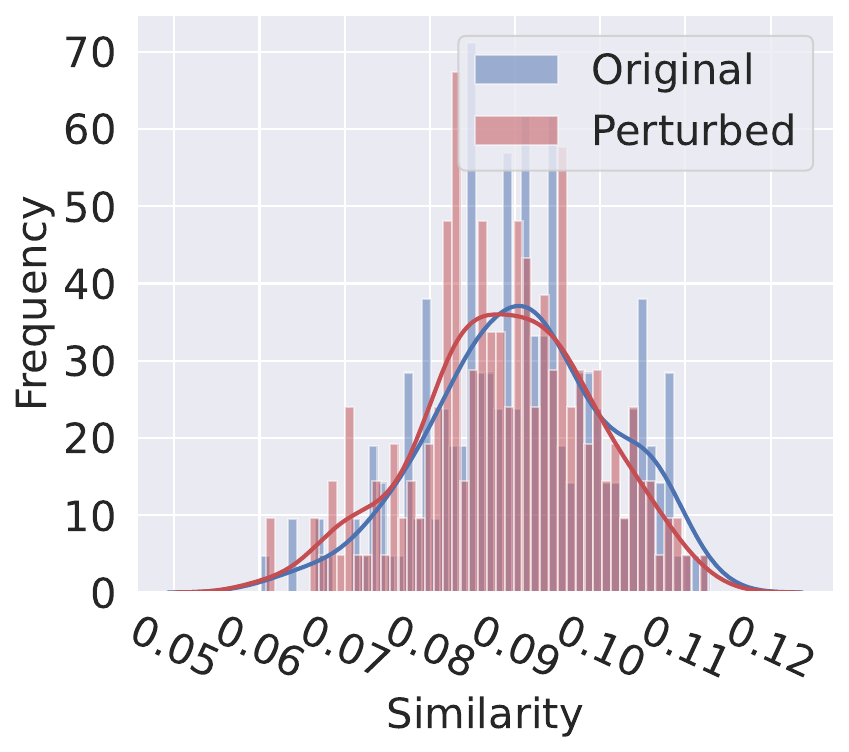}
    \caption{JSM-inf}
    \label{fig:jsm-distancemean}
\end{subfigure}
\begin{subfigure}{0.3\linewidth}
\centering
    \includegraphics[width=\linewidth]{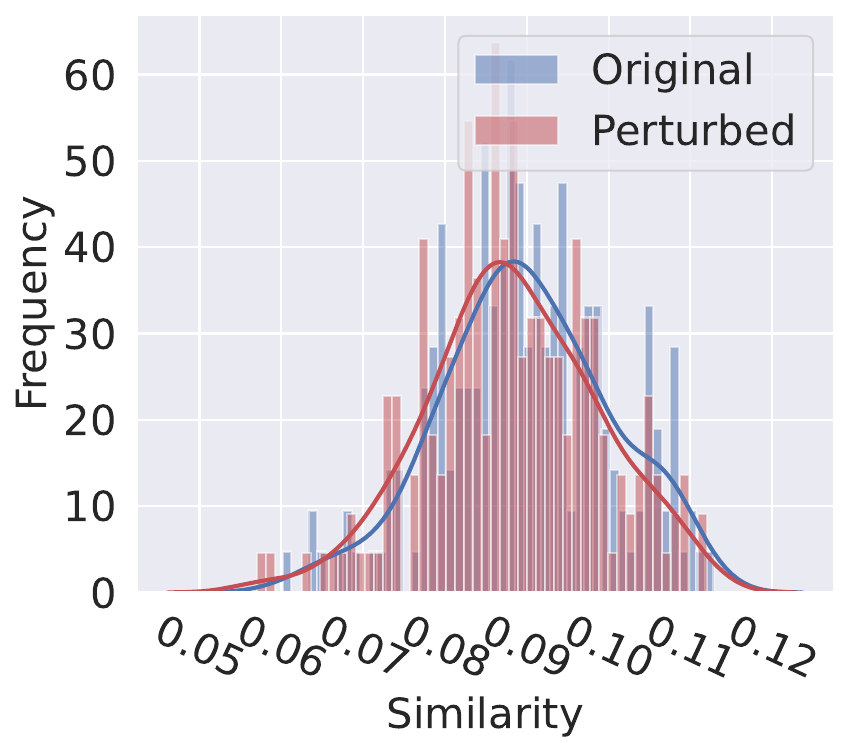}
    \caption{DF-inf}
    \label{fig:df-distancemean}
\end{subfigure}
\begin{subfigure}{0.3\linewidth}
\centering
\includegraphics[width=\linewidth]{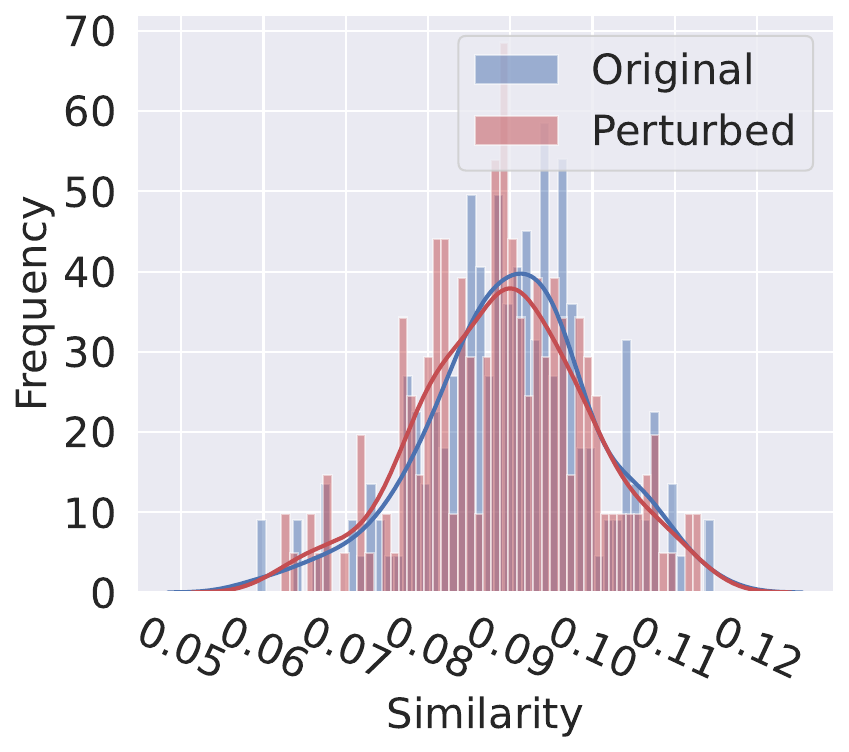}
    \caption{Pixel-3}
    \label{fig:pix-distancemean}
\end{subfigure}
\begin{subfigure}{0.3\linewidth}
\centering
    \includegraphics[width=\linewidth]{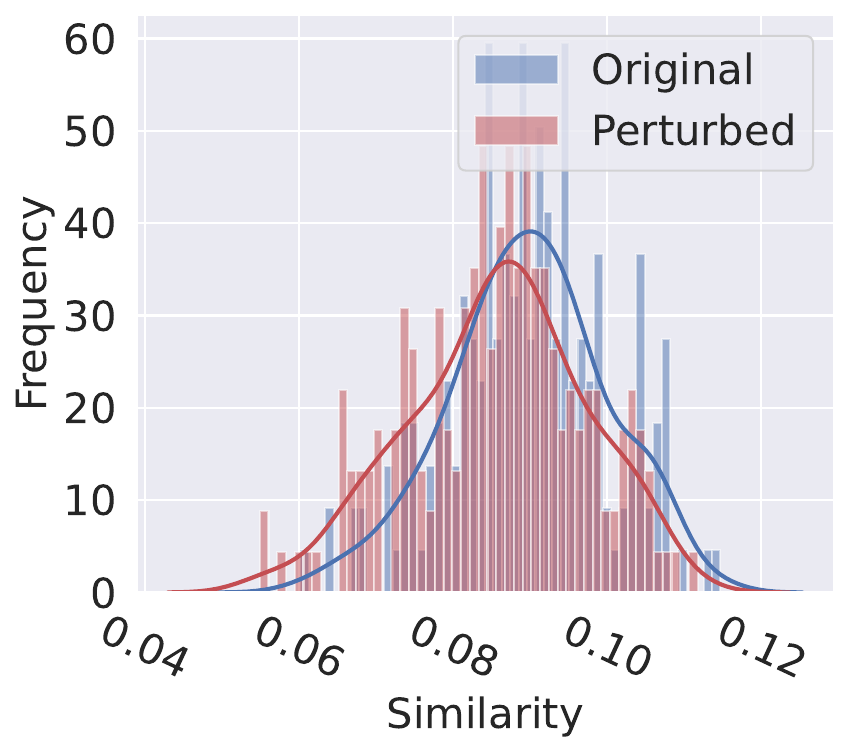}
    \caption{C\&W-inf}
\label{fig:cwi-distancevar}
\end{subfigure}
\caption{Sample-wise linear feature similarly distributions of original and adversarial samples with different adversarial attacks for CIFAR-10, demonstrating the relative differences.}
\label{fig:feature_sim_uni}
\end{minipage}
\hfill
\begin{minipage}{0.30\linewidth}
\begin{subfigure}{0.99\linewidth}
\centering
\begin{subfigure}{0.32\linewidth}
\includegraphics[width=\linewidth]{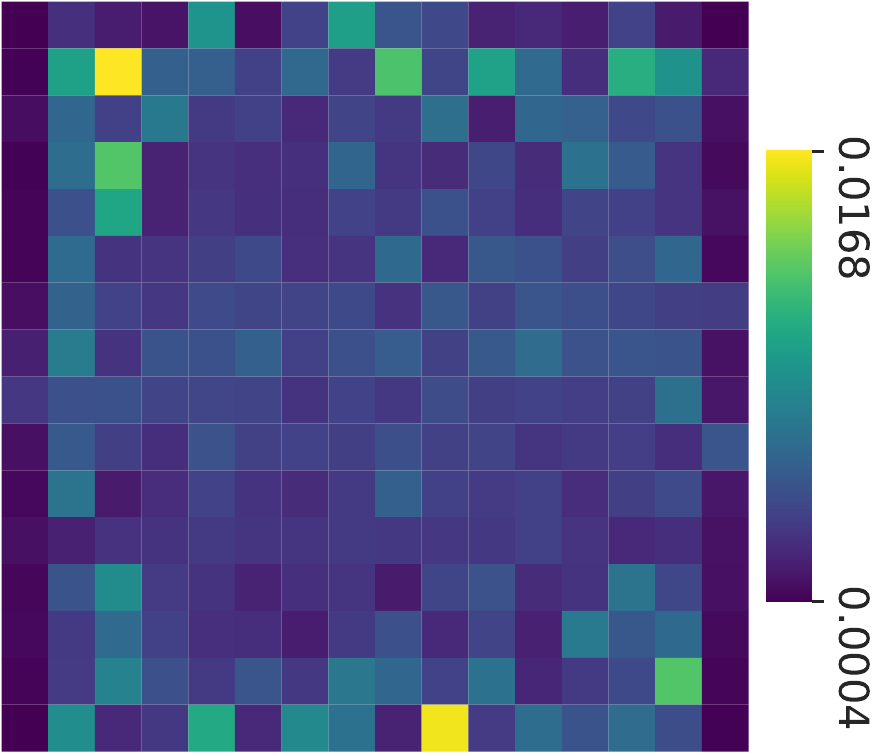}
\end{subfigure}
\begin{subfigure}{0.32\linewidth}
\includegraphics[width=\linewidth]{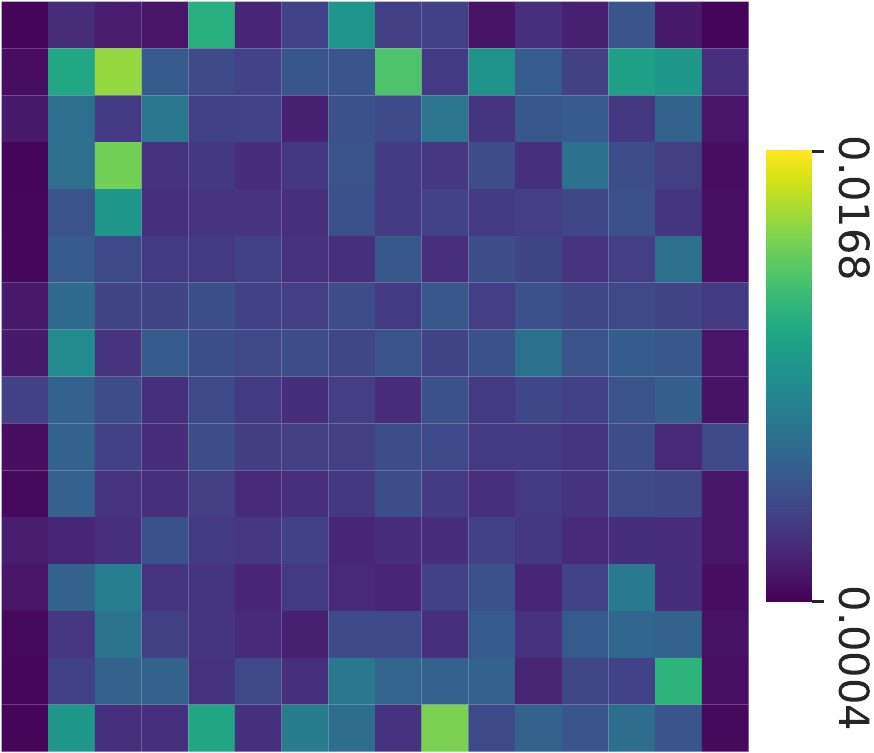}
\end{subfigure}
\begin{subfigure}{0.32\linewidth}
\includegraphics[width=\linewidth]{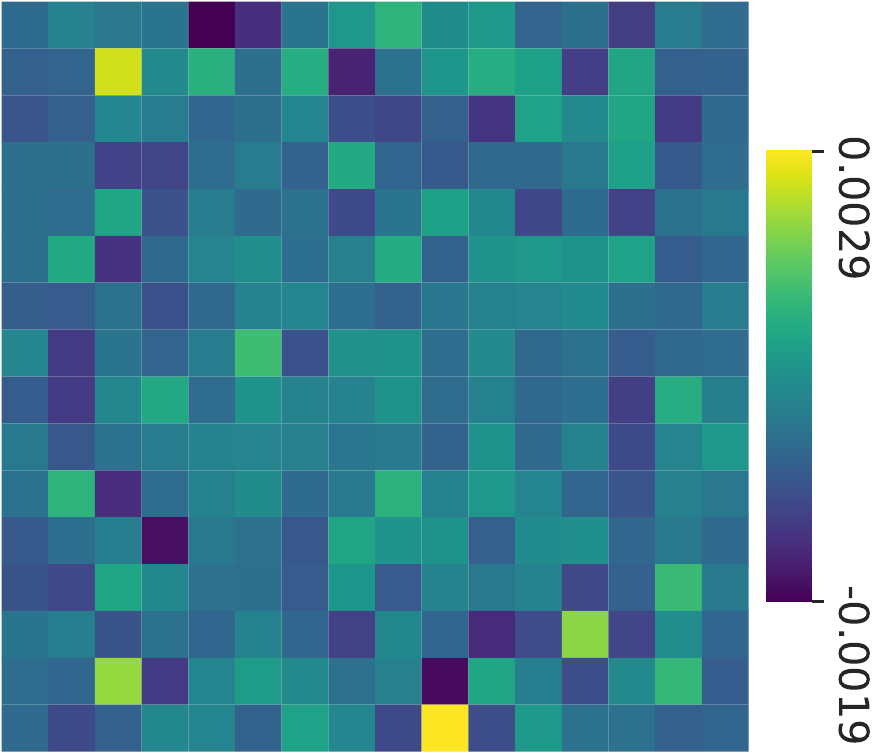}
\end{subfigure}
\caption{Org/Adv/Diff (Cross-attention)}
\label{fig:att_security}
\end{subfigure}
\begin{subfigure}{0.99\linewidth}
\centering
\begin{subfigure}{0.32\linewidth}
\includegraphics[width=\linewidth]{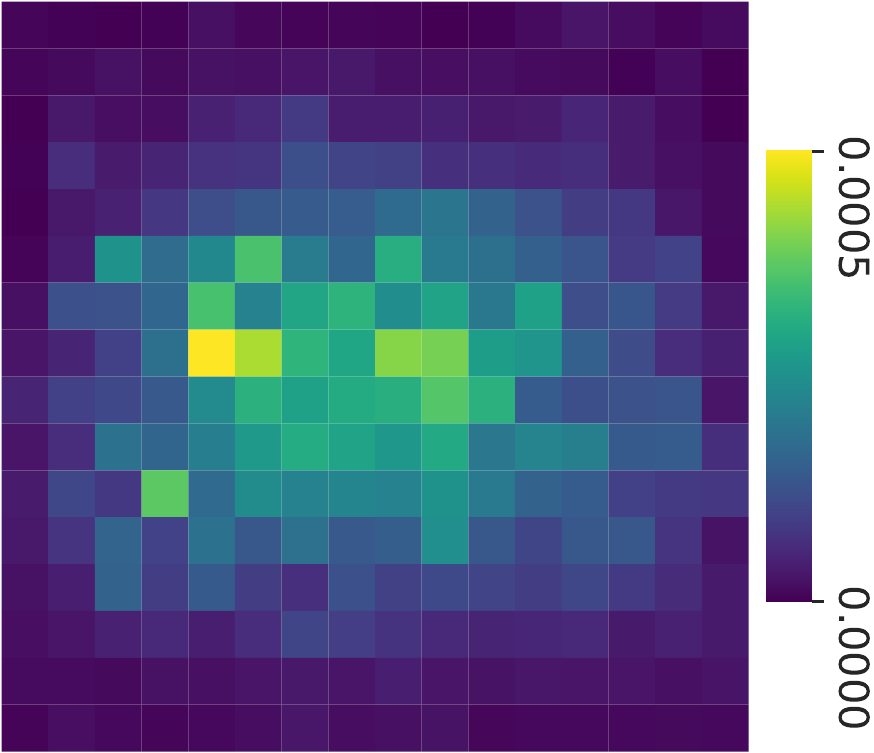}
\end{subfigure}
\begin{subfigure}{0.32\linewidth}
\includegraphics[width=\linewidth]{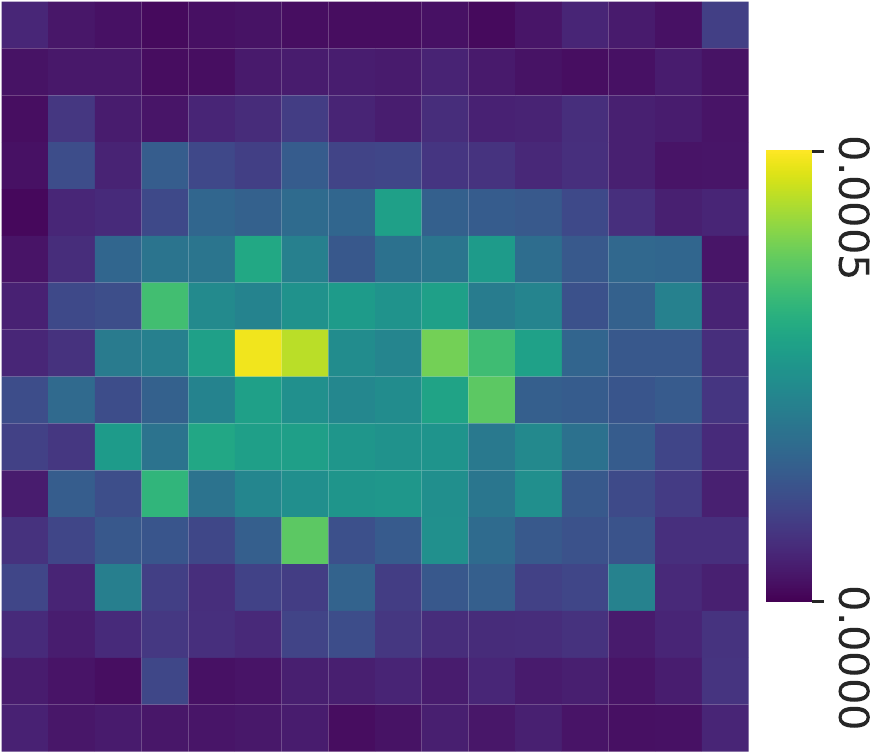}
\end{subfigure}
\begin{subfigure}{0.32\linewidth}
\includegraphics[width=\linewidth]{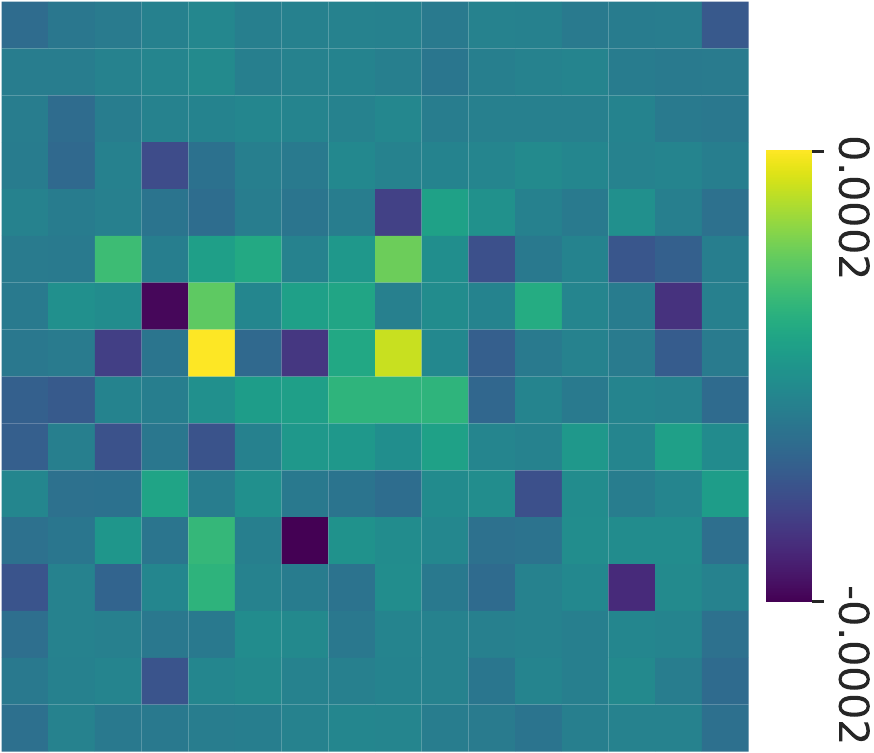}
\end{subfigure}
\caption{Org/Adv/Diff (Grad-CAM)}
\label{fig:grad_security}
\end{subfigure}
\caption{Original, adversarial and difference between of prominent concept cross-attention maps and Grad-CAM on attention maps with C\&W attack for CIFAR-10. 
}
\end{minipage}
\hfill
\begin{minipage}{0.33\linewidth}
    \centering
    \includegraphics[width=0.93\linewidth]{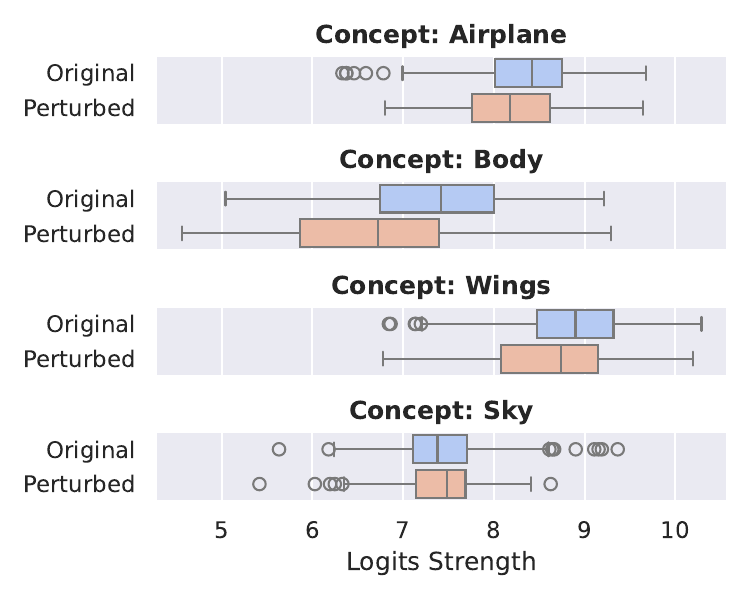}
    \caption{Concept posterior strength distributions of concepts for CIFAR-10 with a FGSM-16 attack transferring samples from airplane to bird.}
    \label{fig:concept_dependency}
\end{minipage}
\end{figure*}

\noindent \textbf{Model resistance to perturbations.}
We use the unimodal model settings on CIFAR-10 to evaluate, and assume that the model provider collects adversarial samples labeled with the attacked class and manually reassigns them to their correct labels. The first investigation, which explores the distribution of the sample-aware linear feature similarity between original and various attack datasets, is depicted in Figure~\ref{fig:feature_sim_uni}. Figures~\ref{fig:fgsm-distancemean} and~\ref{fig:pgd-distancemean}, show how attacks with obvious perturbations (\eg FGSM-16/255 and PGD-16/255) will demonstrate distinct histogram patterns with clearly separable peaks, making them a valid candidate for classification by threshold. However, when it comes to minimal few-pixel perturbations (Figure~\ref{fig:pix-distancemean}, which shows a 3-pixel change attack), the distance distributions overlap, and it becomes challenging to separate attack samples from original samples based solely on the similarity distribution. \textit{This finding reveals the underlying mechanism of perturbation-based attacks}: In cases of larger perturbations, adversarial attacks can alter the semantics of the image. Conversely, even with minimal perturbations that do not significantly change the image's semantics, the model can still be effectively perturbed, exploiting the model's inherent weaknesses. 

Given that C\&W attacks are designed by optimizing a target function to minimize the perturbation of adversarial samples ~\cite{carlini2017towards}, we can use the semantic shift between successful C\&W attack samples and original samples as a scoring mechanism for evaluating the model's \textit{resistance} to perturbations.
We therefore investigated the attention mechanisms of adversarial samples generated by C\&W attacks, focusing on key concept words $Concept_\text{pro}$ to identify the regions of the input samples that are most susceptible to perturbation. Comparing the cross-attention maps  extracted from both original and adversarial samples gives us the differences depicted in Figure~\ref{fig:att_security}.

The heatmaps provide a spatial representation of the changes between the original and attacked samples. Under minimal perturbation, the proportion of positions where attention shifts is small. This suggests that while minimal perturbations introduce discrepancies, these are localized to specific regions of the image rather than being dispersed across the entire map. These regions can therefore serve as focal points for designing defenses that minimize the impact of such perturbations. The gradient maps in Figure~\ref{fig:grad_security} are derived from the cross-attention maps of both the original and C\&W attack adversarial samples and show regions with significant differences between the original and adversarial samples.

\noindent \textbf{Model conceptual vulnerabilities.}
Next, we focus on analyzing a specific type of misclassified sample: instances incorrectly categorized from Class A to Class B. By observing how these erroneous samples shift the concepts inherent to Class A, we can infer the model's vulnerabilities when learning concepts from Class A. Specifically, we analyze adversarial samples generated using FGSM-16, where instances from the ``Airplane'' class are misclassified as the ``Bird'' class in the CIFAR-10 dataset. This focus is motivated by the previous experiment in Section~\ref{sec:uni_resistance}, where FGSM-16 was found to effectively perturb the semantic concepts of samples. By examining the specific concepts that are disrupted, we can determine the model's dependency on those concepts.

Figure~\ref{fig:concept_dependency} illustrates the intensity of different concepts for both the original and adversarial samples within this subset. It is evident that for the prominent concept word ``Airplane'', a subtle shift occurs, indicating that the adversarial sample has indeed affected the semantic understanding of ``Airplane''. However, for the concept words ``Body'' and ``Sky'', there is a more significant divergence, suggesting that the target model's classification of the ``Airplane'' category relies heavily on these two concepts.
Consequently, when perturbations alter the input's relationship to these concepts, the model's classification process is disrupted.

\begin{figure}[t]
\centering
\includegraphics[width=\linewidth]{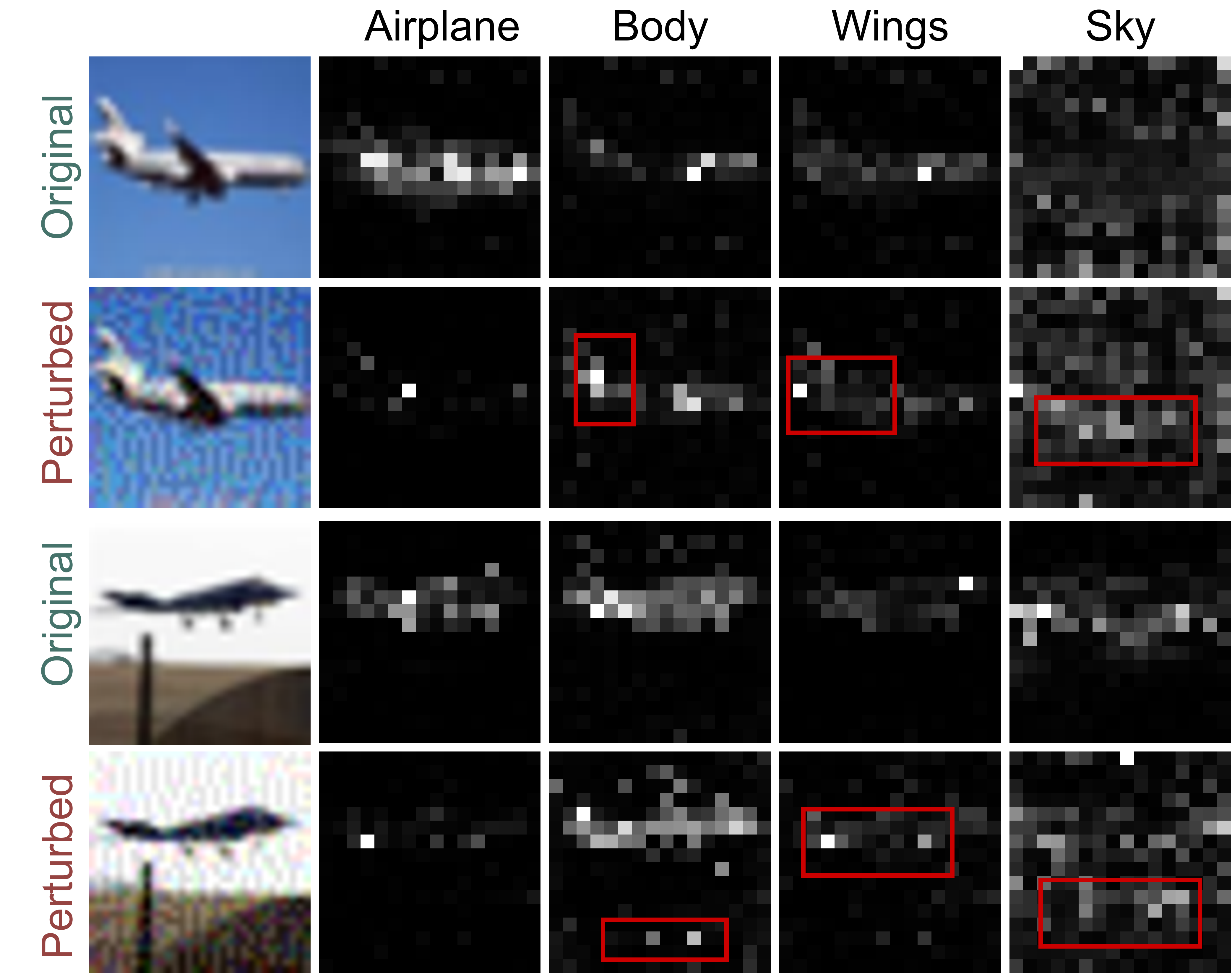}
\caption{Two examples illustrate the difference in Grad-CAM attention for various concept words extracted from original and adversarial samples in CIFAR-10 under the FGSM-16 attack. These examples, which involve transferring samples from airplanes to birds, demonstrate the conceptual-level perturbations affecting model decisions.}   
\label{fig:simple_wise} 
\vspace{-1em}
\end{figure}

For sample-wise analysis, we performed Grad-CAM visualizations for each relevant concept. 
Figure~\ref{fig:simple_wise} depicts that the attack effectively diminishes the attention on the concept word ``Airplane'' in both samples. 
For the first example, the position of the ``wings'' is misinterpreted, making the area resemble a bird spreading its ``wings''. Additionally, the model's attention to the ``sky'' is intensified, creating a connection with the main object and leading the model to classify it as a bird. In the second example, the attention on the ``body'' is dispersed throughout the image, leading the model to potentially perceive the scene as consisting of one large bird and two smaller birds. Simultaneously, the ``sky'' extends to the lower part of the image, further contributing to the misclassification. This reveals the model's instability when these specific concepts are perturbed.

\begin{takeaway}
\addtocounter{takeaway}{1}
\noindent\textbf{Takeaway \thetakeaway: }
\textit{ \tool{} identified low-perturbation attacks (e.g. C\&W attack) that exploit the model's weaknesses in specific input regions. 
Moreover, \tool{} identified concepts the target model is overly dependent on, which suspicious concepts are misleading it, and how disrupting key concepts negatively impacts the model. These weaknesses in concepts arise from the model's inability to fully learn all the concepts associated with each class from the training data.}
\end{takeaway}

\subsection{Model Bias}\label{sec:meme}

In this section, we will explore whether \tool{} can detect shifts in samples related to biased concepts and thereby identify the locations in the image associated with biased semantics. 
While models may exhibit various forms of bias (e.g., gender bias, racial bias), this study specifically focuses on sociological bias due to its subtlety. However, the proposed methodology can be adapted to address other types of bias by evaluating different concepts.
Previous works~\cite{qu2023evolution,qu2023unsafe} manually quantified image bias, yet \tool{} will allow us to automatically quantify the extent to which a sample exhibits sociological bias. 

In image generation models, users may deliberately craft prompts to induce biased outputs. To preserve utility, the model may be compelled to generate images containing sociological bias. We first investigate whether \tool{} is capable of detecting sociological symbols embedded (\ie sociological bias) in the generated images. Furthermore, the model's behavior may also vary depending on the embedded sociological symbols. For example, it may tend to generate high-quality toxic memes more frequently when specific country flags are used compared to others. Leveraging \tool{}'s ability to capture sociological symbols, we further assess such model bias by evaluating the quality of generated samples associated with different societies.

Different from the data bias injection discussed in Section~\ref{sec:data-bias}, where we evaluate the model's implicit preferences when no specific concept is provided (\eg using the prompt ``a girl drinking cola'', and observing whether the model tends to generate Coca-Cola), here we evaluate the model’s ability to generate explicitly specified concepts. For example, we prompt the model to generate specific country flags in memes and evaluate its generation quality for each, in order to interpret the model’s bias toward different entities.

\vspace{1mm}
{\color{red}
\noindent \textbf{Note: This next section includes some discriminatory content that may offend some readers. We have chosen to show these examples for illustration only and to arouse public awareness of this potential risk. }}

\begin{figure}[h]
\centering
\includegraphics[width=\linewidth]{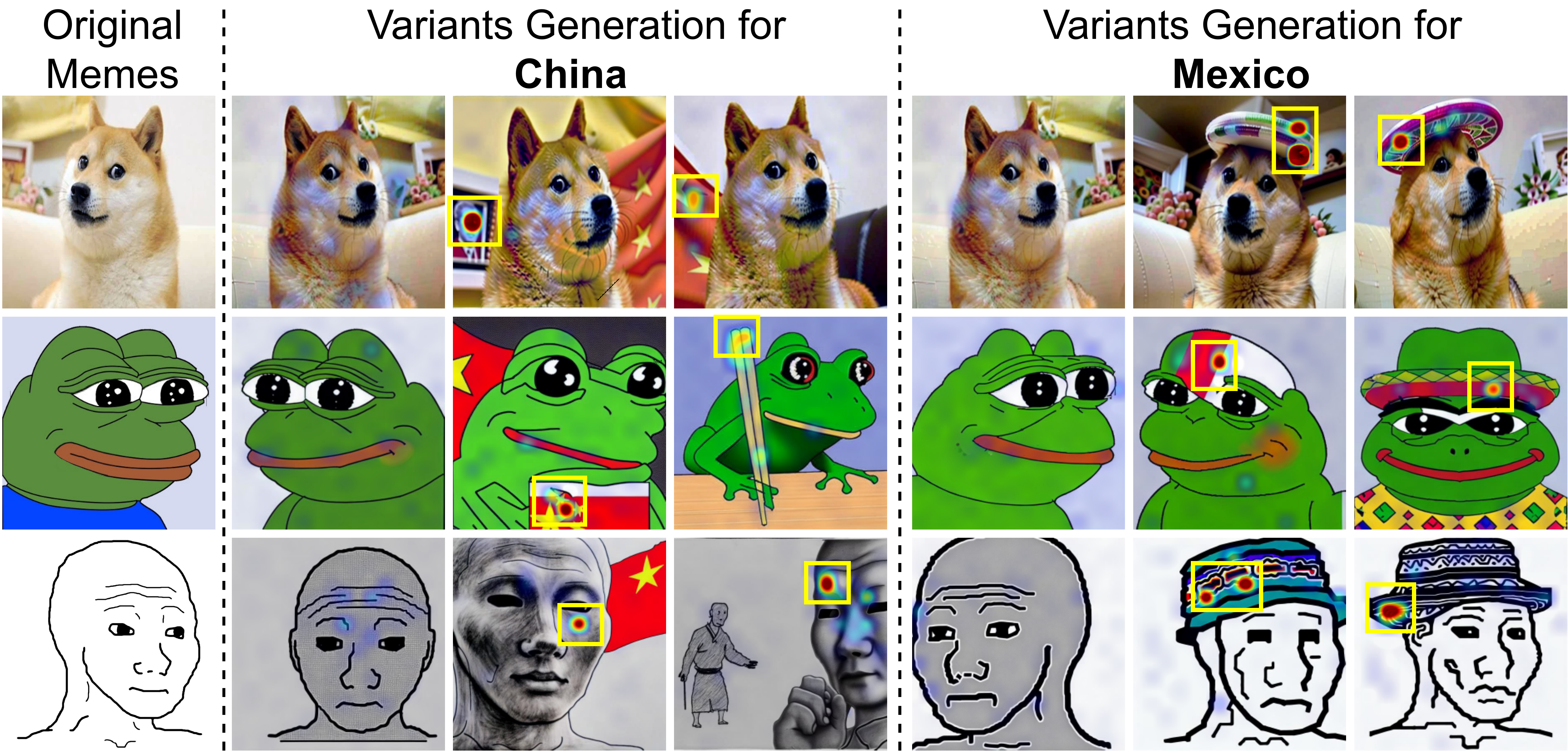}
\caption{Grad-CAM attention overlaid on memes representing different stereotypes verifies the ability of \tool{} to investigate bias. High-attention areas are highlighted using a yellow frame.}
\label{fig:meme} 
\vspace{-1.5em}
\end{figure}

\vspace{-0.7em}
\subsubsection{Integrity Evaluation: Bias Localization} 

\noindent \textbf{Image editing model.} Following prior work that used text-to-image models to generate biased memes~\cite{qu2023unsafe}, we employ DreamBooth~\cite{ruiz2023dreambooth} as a learning-based technique designed for image editing, which has been previously identified as a potential method for biased image generation.

\noindent \textbf{Generation process for bias quantification.} For a given image sample, sociological associations are often evoked by the presence of distinctive sociological symbols (\eg objects, clothing, style, color schemes). By generating samples that explicitly include such sociological symbols, we evaluate whether \tool{} can successfully associate these features with the corresponding society, thereby identifying samples that evoke sociological associations in human perception.

To achieve this, we selected three widely recognized memes that lack any inherently biased connotations -- namely, ``Doge'', ``Pepe the Frog'', and ``Wojak'' to generate variants. Using ChatGPT, we created stereotype keywords related to ``Mexicans'' and ``Chinese'' to generate meme variants with biases towards these groups. For analysis, we focused on the most visually apparent samples.

We first overlay the heatmap by normalizing the Grad-CAM attention focused on the country concept onto the generated memes to observe the intensity and location of the country-related concept within the image. Figure~\ref{fig:meme} shows our \tool{} framework identifying biased regions across variants of the three different memes, with attention consistently focused on items and elements with biased connotations. Additionally, for unsuccessful generations (as seen in the first column), the attention does not concentrate, indicating a lower tendency to flag benign samples as biased.

We found that the attention mechanism successfully identifies unique sociological symbols across different cultures. For example, in the second column,  \tool{} associates the depicted chopsticks as a symbol of China, a reasonable if stereotypical link. This demonstrates \tool{}'s ability to discern implicit sociological features and bias, allowing it to quantify the relationship between generated content and a specific society.

\subsubsection{Attribution: Models' Generation Ability} Based on these insights for overlapping the Grad-CAM, we designed a new quantification to attribute models' biased generation across different society in Appendix~\ref{app:bs}. An ideal, unbiased generative model should exhibit uniform generation quality across all societies. We designed two case studies, and investigated that current meme generation models exhibit consistent behavior across different society. 
\begin{takeaway}
\addtocounter{takeaway}{1}
\noindent\textbf{Takeaway \thetakeaway: }
\textit{\tool{} effectively locates sociological bias within memes and provides a heuristic for the degree of bias present, providing valuable insights into how sociological biases are reflected and propagated in generated content. This capability allows us to assess the performance of AIGC models in memes editing across different society, where we observed significant variations. The generation performance is stronger for certain countries, likely due to their sociological prominence causing the unbalanced training samples. This could be used to reveal inherent biases and unfairness in current generative AI.}
\end{takeaway}

\begin{takeaway}
    \addtocounter{takeaway}{1}
    \noindent\textbf{Takeaway \thetakeaway: }
    \textit{While adversarial robustness and unreliable generation are model-level issues, they are still rooted in deficiencies at the data level. The imbalance and lack of comprehensiveness in training data concepts result in the model's insufficient understanding of certain concepts. Additionally, the unequal representation of concepts across different groups in the training data leads to biases in the model.}
\end{takeaway}

\section{Conclusion}

The paper introduces \tool{} as a novel approach to understanding and addressing integrity in AI models during training and inference by analyzing conceptual shifts, effectively tackling both intentional and unintentional risks. 
\tool{} demonstrates strong detection performance against vanilla poisoning attacks while uncovering a new bias injection threat driven by malicious concept shifts, such as covert advertisements. It identifies unintentional samples with privacy risks and evaluates the influence of specific concepts on model memorization. For model-level risks, \tool{} achieves high anomaly detection rates for image classifiers and multimodal VLP attacks, revealing overreliance on certain concepts and the negative impact of disrupting key concepts, emphasizing the need for comprehensive concept learning. Additionally, \tool{} quantifies sociological bias in AIGC models, particularly memes, by localizing and measuring biases tied to sociological symbols. 
We attribute model-level issues to the lack of integrity and balance in the representation of concepts within the training data.

We acknowledge that \tool{} can only identify concepts on which a target model is overly dependent, and this may not always be sufficient to understand the root cause of a failure. Causality analysis through concept shift is one future work. The paper also does not explore the dynamic nature of concepts. Another avenue for future work is to study the semantic meaning of a concept which can evolve over time, influenced by societal factors.

\section*{Acknowledgments}

Minhui Xue is supported in part by Australian Research Council (ARC) DP240103068 and in part by CSIRO – National Science Foundation (US) AI Research Collaboration Program. Hammond Pearce is supported in part by ARC DP250101396.

\bibliographystyle{ACM-Reference-Format}

\bibliography{ref}

\appendix
\section*{Appendix}

\section{Ethical Considerations}
Our work aims to evaluate the risks in deep learning models from the model training stage to the inferencing stage. We do this by building a multimodal analysis framework ~\sol{}. There are no potential harms associated with our research. We use this to analyze the trustworthy risks faced by critical models following Figure~\ref{fig:flow} and use our framework to examine integrity anomaly detection for probing samples, attributing models' faults.

\section{Open Science}
The code, extended results and artifacts are made available at GitHub: 
\url{https://github.com/trust-in-ai/conceptlens}

\section{Bias score}
\label{app:bs}
Based on these insights for overlapping the Grad-Cam, we developed a pair $(s, g)$ for a specific society, where the first component $s$ is a  score by combining the linear similarity defined in \Circled{F1} in Section~\ref{sec:feature_extract} and the second component denotes the maximum value in a matrix of Grad-CAM attention  $g = \max_{i,j} G_{i,j}$, where the matrix $G$ is defined in \Circled{F3} in Section~\ref{sec:feature_extract}. This score is used to rank all generated samples. We finally stretched the pair $(s, g)$ and normalized it into obtaining an ultimate bias score for a specific society as follows: $\text{Bias Score} = \alpha \times \frac{s - \min(\textbf{s})}{\max(\textbf{s}) - \min(\textbf{s})} + \alpha  \times \frac{g - \min(\textbf{g})}{\max({\textbf{g})} - \min(\textbf{g})},$ where $(\textbf{s}, \textbf{g})$ belongs to all sample pairs, $\alpha$ as a constant (here we use 0.5). It is worth noting that one limitation of this score is that both the Grad-CAM attention intensity and the bias score are relative measures, which only allow for comparisons within each individual cultural group, rather than across different cultures.

\end{document}